\PassOptionsToPackage{unicode}{hyperref}
\PassOptionsToPackage{hyphens}{url}
\documentclass[
  11pt,
]{article}
\usepackage{amsmath,amssymb}
\usepackage{lmodern}
\usepackage{iftex}
\ifPDFTeX
  \usepackage[T1]{fontenc}
  \usepackage[utf8]{inputenc}
  \usepackage{textcomp} 
\else 
  \usepackage{unicode-math}
  \defaultfontfeatures{Scale=MatchLowercase}
  \defaultfontfeatures[\rmfamily]{Ligatures=TeX,Scale=1}
\fi
\IfFileExists{upquote.sty}{\usepackage{upquote}}{}
\IfFileExists{microtype.sty}{
  \usepackage[]{microtype}
  \UseMicrotypeSet[protrusion]{basicmath} 
}{}
\makeatletter
\@ifundefined{KOMAClassName}{
  \IfFileExists{parskip.sty}{%
    \usepackage{parskip}
  }{
    \setlength{\parindent}{0pt}
    \setlength{\parskip}{6pt plus 2pt minus 1pt}}
}{
  \KOMAoptions{parskip=half}}
\makeatother
\usepackage{xcolor}
\usepackage[margin=1in]{geometry}
\usepackage{graphicx}
\makeatletter
\def\maxwidth{\ifdim\Gin@nat@width>\linewidth\linewidth\else\Gin@nat@width\fi}
\def\maxheight{\ifdim\Gin@nat@height>\textheight\textheight\else\Gin@nat@height\fi}
\makeatother
\setkeys{Gin}{width=\maxwidth,height=\maxheight,keepaspectratio}
\makeatletter
\def\fps@figure{htbp}
\makeatother
\newlength{\cslhangindent}
\newlength{\csllabelwidth}
\newlength{\cslentryspacingunit} 
\newenvironment{CSLReferences}[2] 
 {
  \setlength{\parindent}{0pt}
  \ifodd #1
  \let\oldpar\par
  \def\par{\hangindent=\cslhangindent\oldpar}
  \fi
  \setlength{\parskip}{#2\cslentryspacingunit}
 }%
 {}
\usepackage{calc}

\usepackage{amsmath,amssymb}
\numberwithin{equation}{subsection}
\usepackage{lineno}

\usepackage{float}
\usepackage{setspace}
\onehalfspace
\ifLuaTeX
  \usepackage{selnolig}  
\fi
\IfFileExists{bookmark.sty}{\usepackage{bookmark}}{\usepackage{hyperref}}
\IfFileExists{xurl.sty}{\usepackage{xurl}}{} 
\hypersetup{
  pdftitle={Topical Hidden Genome: Discovering Latent Cancer Mutational Topics using a Bayesian Multilevel Context-learning Approach},
  pdfauthor={Saptarshi Chakraborty; Zoe Guan; Colin B. Begg; Ronglai Shen},
  pdfkeywords={multilevel Bayesian models;, whole genome data,, rare
somatic variants,, topic model,, context learning,, Markov chain Monte
Carlo},
  hidelinks,
  pdfcreator={LaTeX via pandoc}}

\title{Topical Hidden Genome: Discovering Latent Cancer Mutational
Topics using a Bayesian Multilevel Context-learning Approach}
\author{Saptarshi Chakraborty\footnote{Department of Biostatistics,
  State University of New York at Buffalo,
  \href{mailto:chakrab2@buffalo.edu}{\nolinkurl{chakrab2@buffalo.edu}}} \and Zoe
Guan\footnote{Department of Epidemiology \& Biostatistics, Memorial
  Sloan-Kettering Cancer Center,
  \href{mailto:guanz@mskcc.org}{\nolinkurl{guanz@mskcc.org}}} \and Colin
B. Begg\footnote{Department of Epidemiology \& Biostatistics, Memorial
  Sloan-Kettering Cancer Center,
  \href{mailto:beggc@mskcc.org}{\nolinkurl{beggc@mskcc.org}}} \and Ronglai
Shen\footnote{Department of Epidemiology \& Biostatistics, Memorial
  Sloan-Kettering Cancer Center,
  \href{mailto:shenr@mskcc.org}{\nolinkurl{shenr@mskcc.org}}}}
\date{}

\begin{document}
\maketitle
\begin{abstract}
Statistical inference on the cancer-site specificities of collective
ultra-rare whole genome somatic mutations is an open problem.
Traditional statistical methods cannot handle whole-genome mutation data
due to their ultra-high-dimensionality and extreme data sparsity --
e.g., \textgreater30 million unique variants are observed in the
\textasciitilde1700 whole-genome tumor dataset considered herein, of
which \textgreater99\% variants are encountered only once. To harness
information in these rare variants we have recently proposed the
``hidden genome model'', a formal multilevel multi-logistic model that
mines information in ultra-rare somatic variants to characterize tumor
types. The model condenses signals in rare variants through a
hierarchical layer leveraging contexts of individual mutations. The
model is currently implemented using consistent, scalable point
estimation techniques that can handle 10s of millions of variants
detected across thousands of tumors. Our recent publications have
evidenced its impressive accuracy and attributability at scale. However,
principled statistical inference from the model is infeasible due to the
volume, correlation, and non-interpretability of the mutation contexts.
In this paper we propose a novel framework that leverages topic models
from the field of computational linguistics to induce an
\emph{interpretable dimension reduction} of the mutation contexts used
in the model. The proposed model is implemented using an efficient MCMC
algorithm that permits rigorous full Bayesian inference at a scale that
is orders of magnitude beyond the capability of out-of-the-box
high-dimensional multi-class regression methods and software. We employ
our model on the Pan Cancer Analysis of Whole Genomes (PCAWG) dataset,
and our results reveal interesting novel insights.
\end{abstract}

\allowdisplaybreaks

\newcommand{\N}{\operatorname{N}}

\newcommand{\Ga}{\operatorname{Gamma}}

\newcommand{\Poi}{\operatorname{Poisson}}

\newcommand{\appropto}{\mathpalette\approptoinn\relax}

\newcommand{\ds}{\displaystyle}

\newcommand{\Wt}{\tilde{W}}

\newcommand{\Ht}{\tilde{H}}

\hypertarget{intro}{%
\section{Introduction}\label{intro}}

A large and growing body of research has documented strong links between
specific somatic mutations and cancer types (Haigis, Cichowski, and
Elledge 2019). This has put forward an emerging field aiming to
characterize cancer types through tumor mutation footprints. Some
notable clinical applications are as follows: (a) predicting primary
sites of origin of metastasized tumors of uncertain primaries, a major
clinical task involving 3-5\% of all cancer diagnoses worldwide; (b)
accurate prognostication of rare, phenotypically heterogeneous cancer,
and (c) early detection of cancer from liquid biopsies through
circulating tumor DNAs (ctDNAs) in the bloodstream. The mutation driven
cancer-site characterization framework may provide rigorous statistical
answers to these pertinent questions.

The existing research in this area has two broad ends. On the
\emph{clinical} end lie studies considering prediction of tumors based
on their mutation footprints (Soh et al. 2017; Jiao et al. 2020). These
studies train classification models, typically blackbox methods such as
deep learning, support vector machine, random forest etc., on somatic
mutation data and then use the trained models to predict cancer types of
new tumors. These ``pure prediction algorithms'' permit high predictive
accuracy in specific examples, but they lack interpretation, attribution
(Efron 2020) and inference, and may produce results that are unreliable
for clinical tasks (Lapuschkin et al. 2019).

By contrast, on the \emph{scientific end} lie studies investigating and
quantifying associations between specific mutations and cancer to derive
a curated list of specific genomic regions, e.g., of cancer genes
(Chakravarty et al. 2017), and subsequently utilizing this curated list
to understand the biology of cancer risk (Soh et al. 2017; Haigis,
Cichowski, and Elledge 2019). These studies typically focus on a handful
of recurring variants/genomic regions to infer their tissue
site-specificities via some established inferential methods, but ignore
the rare and previously unobserved variants. This, in part, is due to
the lack of statistical methods available for handling ultra-high
dimensional and rare somatic mutations. Indeed, the vast majority of
whole genome alterations are ultra-sparse -- e.g., \textless1\% of the
\textgreater35 million of variants cataloged in the PCAWG data (Campbell
et al. 2020) considered herein comprising \textgreater1500 tumors were
detected in more than one tumor. Novel mutations are also routinely
discovered in new sequencing studies. Such data are not suited for
traditional inferential statistical methods. However, when harnessed via
appropriate signal condensation tools such as the \emph{Good-Turing}
empirical Bayes probability estimates, strong tissue-specific
information can be extracted from rare variants, as demonstrated in our
earlier empirical study (Chakraborty et al. 2019).

Within this framework, we have recently proposed the \emph{hidden genome
model} (Chakraborty, Begg, and Shen 2020)\emph{,} a formal multilevel
multinomial logistic model that characterizes cancer types by using the
tumor's entire somatic mutation fingerprint including the ultra-sparse
variants as predictors. To harness this ultra-massive predictor space
the model employs \emph{context-based feature condensation}. More
precisely, its employs in a hierarchical layer \emph{meta-features}
describing individual variants -- such as gene labels, types of single
base substitution (SBS; 96 possible types), topological positions on the
chromosome, and various epigenetic features -- to model the regression
effects of the individual variants themselves (see Section
\ref{sec:hidgen-model}). This contextualization is based on an intuitive
observation: for the vast majority of individual variants, including all
rare variants, mutation contexts can describe collectively the entirety
of their tissue-specific effects, with only a handful of frequent
variants exhibiting strong residual effects that are not explained by
their contexts alone. This contextualization aids highly informative
predictor dimension reduction -- condensing information in 10s of
millions of individual variants into a few thousand meta-features. This,
together with appropriately defined sparsity inducing layers in the
model effectuates stable, consistent, and highly scalable (marginal
maximum a posteriori) point estimation using group-lasso (Yuan and Lin
2006) regularized multi-logit likelihood maximization. The model has
been successfully employed on whole genome data with \textgreater20
million variants (Chakraborty, Martin, et al. 2021), and also in
targeted gene-panel based clinical cancer research setting (Chakraborty,
Ecker, et al. 2021) and has provided rigorous quantification of
tissue-specificities of collective mutations while permitting impressive
prediction/classification accuracy in each application.

However, quantifying estimation uncertainty for statistically stable and
biologically meaningful inference is currently infeasible for two
primary reasons. First, the parameters associated with the meta-feature
categories, while aiding substantial dimension reduction for point
estimation, are still too voluminous (in the order of
\(K \times 1000\)s, \(K\) being the number of cancer categories) for
MCMC-based full Bayesian implementation of the multi-level model.
Second, the correlatedness and lack of biological interpretability of
individual mutation context categories such as the \emph{observed} 96
SBS types and the windowed positions on the chromosomes, preclude
scientifically meaningful inference.

In this paper, we posit a novel methodological framework that builds
upon this previous work, introducing new key structures that
collectively permit \emph{interpretable} and \emph{practicable} (full)
Bayesian inference at scale. The framework proposes \emph{interpretable
dimension reduction} of meta-feature categories leveraging topic models
from computational linguistics (Blei and Lafferty 2009). Analogous to
word topics in computational linguistics, which are latent lower
dimensional embeddings that group ``similar'' individual words with
common underlying semantics, meta-feature topics are embeddings of
``similar'' meta-feature categories (e.g., specific genes or specific
chromosome windows); where the ``similarity'' is measured with a view to
tumor type specificities of associated mutations. The meta-feature
topics share important analogies with \emph{mutation signatures} in
genomics, latent structures derived from the observed SBS-96 categories.
We note however that herein we employ a generalized notion of topics
that applies beyond SBS as typically considered in genomics; our
framework caters to any categorical meta-feature source such as genes
and chromosome regions.

In our experience to date the prominent meta-feature topics are
substantially lower dimensional (typically \textless300 in number) and
much less correlated than the original meta-feature categories. This
effectuates impressive bi-level dimension-reduction in the
classification model: first from the variants (10s of millions) to the
observed meta-features (a few thousands) and then from the observed
meta-features to the latent meta-feature topics (typically
\textless300). The regression coefficients of the original ``observed''
meta-features can then be derived as linear combinations of the reduced
meta-feature regression coefficients (c.f. principal component
regression) to quantify observed meta-feature effects, and to also
readily permit full Bayesian posterior prediction from the model on new
datasets. For the substantially reduced-dimensional ``independent''
predictor effects (i.e., latent topic effects and \emph{residual
variant} \emph{effects}, see Section \ref{sec:meta-features}) a Bayesian
group-lasso prior can be elicited for stable estimation and prediction.
Note that both the topic modeling of observed meta-feature categories,
and subsequent hidden genome modeling with meta-feature topics are done
in a single, fully connected multilevel model that permits coherent,
rigorous full Bayesian inference. For implementation, we propose an
efficient Metropolis-within-Gibbs sampler leveraging bi-level data
augmentation Kyung et al. (2010). This permits theoretical guarantees
associated with MCMC methods for accurate posterior inference.

For expository purposes we employ our method on \(1706\) tumors from
\(10\) separate cancer sites sequenced in the Pan Cancer Analysis of
Whole Genomes (PCAWG) study, collectively utilizing \textgreater35
million unique variants as predictors (see Section
\ref{sec:data-description} for a detailed note on the dataset). Our
results reveal interesting novel insights. To assess
interpretability/reproduciblity of our results we validate our findings
through multi-fold followup analyses. For the whole genome chromosome
window topics, in the absence of a direct validation dataset, we
validate our findings through association with certain epigenetic
features. For exome level SBS topics, we perform two direct independent
validations using (a) the publicly reposited COSMIC mutation signatures,
and (b) estimated SBS topics obtained from our model employed on the
Cancer Genome Atlas (TCGA) dataset (subsetted to the same 10 cancer
sites as in the training PCAWG data).

\hypertarget{sec:data-description}{%
\section{Data Description and Problem
Overview}\label{sec:data-description}}

Our primary motivating dataset is from Pan Cancer Analysis of Whole
Genomes (PCAWG) (Campbell et al. 2020). For expository purposes here we
focus on 10 common cancer sites cataloged in the dataset with at least a
moderate sample size (\(n\)) each: liver (\(n\)=349), pancreatic
(\(n\)=326), prostate (\(n\)=286), breast (\(n\)=214), ovarian
(\(n\)=113), kidney (\(n\)=111), skin (\(n\)=107), esophageal
(\(n\)=98), colorectal (\(n\)=60), and lung (\(n\)=38). A total of
36,325,180 variants, with 35,285,233 observed only once, were detected
by a consensus mutation calling approach on these 1,702 tumors: an
average of 19,076 variants per tumor. To understand variant tissue
specificities based on smaller genotyping panels, we consider two
subsetted versions of the whole genome data -- a version to simulate a
whole exome panel (Bailey et al. 2018), and a version to simulate a
targeted cancer gene panel (Zehir et al. 2017). Our objective is to
perform multinomial regression characterizing the cancer type of a tumor
as a function of its mutation footprint (cataloged by the three panels)
while permitting rigorous statistical inference. However, the key
challenge is of course the ultra high-dimensionality and extreme
predictor sparsity. As noted, we propose to solve this problem by
introducing hierarchical layers in the regression model aiding detection
and utilization of latent, lower dimensional meta-feature \emph{topics}.

\hypertarget{sec:model}{%
\section{Full Bayesian Inference on Tissue-Specificities of Whole Genome
Mutations}\label{sec:model}}

\hypertarget{sec:notations}{%
\subsection{Notation and Setup}\label{sec:notations}}

Consider a training dataset comprising \(N\) tumors indexed
\(n=1, \dots, N\) with \(c_n \in \{1, \dots, K\}\) denoting the known
cancer site/class label for tumor \(n\), where \(K\) is the total number
of cancer sites. Label the \(J\) distinct mutations observed in the
training dataset as \(j = 1, \dots, J\); let \(x_{nj}\) be a binary
indicator of the presence of the \(j\)-th mutation in the \(n\)-th tumor
(\(1\equiv\) presence; \(0\equiv\) absence) and let
\(M_n = \sum_{j=1}^J x_{nj}\) be the total mutation burden in tumor
\(n\). Construct the \(N \times J\) variant design matrix \(X\) by
stacking \(x_n = (x_{n1}, \dots, x_{nJ})^T\) as rows:
\(X = ((x_{nj}))\). Thus, \(e_{n; N}^T X = x_n^T\) where \(e_{n; N}\)
denotes the \(N\)-component unit vector whose \(n\)-th component is
\(1\) and the remaining components are \(0\). We are interested in
regressing the cancer site labels \(\{c_n\}\) on the mutation
occurrences \(\{x_{nj}\}\), to (a) jointly infer the tissue-site
specificities of the individual mutations \(\{j\}\), and (b) predict the
cancer label \(c_{n'}\) of a new (test set) tumor
\(n' \notin \{1, \dots, K\}\) through its mutational footprint \(x_n\)
Note that the new tumor \(n'\) may feature hitherto unobserved mutations
\(\{j'\} \notin \{1, \dots, J\}\), and our regression model needs to
appropriately acknowledge this possibility. Below we review the hidden
genome model (Chakraborty, Begg, and Shen 2020) that utilizes
contextualization of variants for this purpose.

\hypertarget{sec:meta-features}{%
\subsection{Contextualization via
meta-features}\label{sec:meta-features}}

The hidden genome model leverages \emph{meta-features,} higher level
features about the individual variants themselves, embodying mutation
contexts. These meta-features are available for all variants, including
rare variants observed in the training data and new variants appearing
only in the prediction (test) data. Genomic meta-features are commonly
categorical, but they may also be continuous. Examples of categorical
meta-features include the gene itself, single base substitution types
(SBS; 96 possible types), and topological position on the chromosome
(e.g., location among \(\sim3000\) 1-megabase length windows). Examples
of continuous meta-features include epigenomic feature scores such as
histone marks and chromatin accessibility. Here we focus primarily on
categorical meta-features to aid topic model based dimension reduction.
Continuous meta-feature variables, if present, can be directly
incorporated without topic modeling into the hidden genome framework
following Chakraborty, Martin, et al. (2021).

To aid distinction we use the term \emph{meta-feature source} to denote
a specific meta-feature variable (e.g., gene, chromosome window) and
\emph{meta-feature categories} to denote individual categories of a
meta-feature source (e.g., gene \emph{KRAS}, chromosome 1 window 1
etc.). Individual meta-feature categories from \emph{all} meta-feature
sources are combined to produce say \(P\) total categories labeled as
\(1, 2, \dots, P\). Subsequently we create dummy coded (``one-hot''
encoded) binary indicators cataloging correspondence between individual
variants and meta-feature categories. For variant \(j\) let
\(u_j = (u_{j1}, \dots, u_{jP})\) be meta-feature indicators with
\(u_{jp}\) being \(1\) if the \(j\)-th variant is associated with
meta-feature category \(p\) and \(0\) otherwise. Continuous meta-feature
values for variant \(j\), if present, are subsequently appended to
\(u_j\). Finally we construct the meta-design matrix \(U\) by stacking
\(u_j\) as rows: \(U = ((u_{jp}))\). The key idea behind the hidden
genome model is to quantify the tissue-specific effects of individual
variants via a meta-feature regression based on the meta-feature design
matrix \(U\).

\hypertarget{sec:hidgen-model}{%
\subsection{A brief review of the point-estimated hidden genome cancer
site classification model}\label{sec:hidgen-model}}

We briefly review below the multilevel \emph{hidden genome model} and
its point estimation as suggested in Chakraborty, Begg, and Shen (2020).
The likelihood layer of the model considers a multi-logistic regression
connecting the cancer type probability \(\Pr(c_n = k)\) to the observed
variant indicators \(\{x_{nj}\}\). A subsequent hierarchical
\emph{meta-feature regression} layer connects variant \(j\) and cancer
site \(k\)-specific regression coefficient \(\beta_{jk}\) to variant
\(j\) specific meta-features \(u_j\) as
\(\beta_{j, k} = u_{j}^T \omega_{\bullet,k} + \beta^0_{j, k}\) where
\(\omega_{\bullet, k} = (\omega_{p, k})_{p=1, \dots, P}\) is the vector
of meta-feature effects specific to cancer site \(k\), and
\(\beta^0_{j, k}\) is the \emph{residual effect} of variant \(j\) in
cancer site \(k\) that is \emph{not} explained by the context. Finally,
a group lasso prior for \(\{\beta_{j,k}^0\}\) and \(\{\omega_{p,k}\}\)
is considered for regularized estimation. The model is succinctly
expressed as follows.

\begin{align}
\Pr(c_n &= k) = \frac{\exp\left[\alpha_k + e_{n;N}^T X\beta_{\bullet, k}\right]}{\exp\left[\sum_{k'=1}^K \alpha_{k'} + e_{n;N}^T X\beta_{\bullet, k'} \right]} \label{eq:model1:layer:mlogit}  \\
\beta_{j, k} &= u_{j}^T \omega_{\bullet,k} +  \beta^0_{j, k} \label{eq:model1:layer:mfeat-regr} \\
\beta^0_{jk} &\sim \operatorname{N}(0, \tau^2_{j;\beta^0}); \
\omega_{pk} \sim \operatorname{N}(0, \tau^2_{p;\omega});  \
\tau^2_{j;\omega}, \tau^2_{p;\theta} \sim \operatorname{Gamma}((K+1)/2, \lambda^2). \label{eq:model1:layer:gr-lasso}
\end{align}

The key meta-feature-regression layer \eqref{eq:model1:layer:mfeat-regr}
regresses \(\{\beta_{jk}\}\) on the contexts \(\{u_j\}\) effectuating
signal condensation in all, including rare, variants. The residual
effects \(\{\beta_{jk}^0\}\) are non-zero and indeed can be
realistically be estimated for only a few strongly discriminative
non-rare variants. Following computation of the \(XU\) matrix arising
after substituting \eqref{eq:model1:layer:mfeat-regr} into
\eqref{eq:model1:layer:mlogit}), one may therefore perform a feature
screening to keep all but a few recurring tissue-specific variants for
estimation of their \(\beta^0_{j,k}\). For scalable point estimation of
the model Chakraborty, Begg, and Shen (2020) marginalize out all but the
key parameters of interest, viz.,
\({{\alpha}}, {{\beta}}^0, {{\omega}}\), given the sparsity level
parameter \(\lambda\) to produce the following log marginal posterior
density. \[
\begin{aligned}
& \log \ \pi({{\alpha}}, {{\beta}}^0, {{\omega}}\mid \lambda, c_1, \dots, c_n, {{x}}_1, \dots, {{x}}_n) \\
&= \sum_{n = 1}^N \sum_{k = 1}^K 1(c_n = k) \log \left( \frac{\exp\left[\alpha_k + e_{n;N}^T X U \omega_{\bullet,k} +  e_{n;N}^T X \beta^0_{j, k} \right]}{\exp\left[\sum_{k'=1}^K \alpha_{k'} + e_{n;N}^T X U \omega_{\bullet,k'} +  e_{n;N}^T X \beta^0_{j, k'} \right]} \right) \\
&\qquad - \lambda \sum_{j = 1}^J \| {{\beta}}_{j, \bullet}^0 \|_2 - \lambda \sum_{p = 1}^P \| {{\omega}}_{p, \bullet} \|_2.
\end{aligned}\] This is a group lasso penalized multi-logistic
log-likelihood with intercept \({{\alpha}}\), and regression
coefficients \({{\beta}}\) and \({{\omega}}\) with associated predictors
\({{x}}_n\) and \({{x}}_n^T{{U}}\) respectively, and penalty parameter
\(\lambda\). The corresponding (marginal) posterior mode can be
efficiently computed using existing software (Friedman, Hastie, and
Tibshirani 2010), and consistency of the resulting estimates follow from
group-lasso theory. Note however that uncertainty quantification for
statistical inference in this framework is infeasible due to
voluminousness of the meta-feature categories.

\hypertarget{sec:full-topic-hidgen-model}{%
\subsection{The topical hidden genome: a Bayesian multilevel model with
interpretable contexts}\label{sec:full-topic-hidgen-model}}

At the outset, we deviate from Chakraborty, Begg, and Shen (2020) by
using instead a normalized version of the predictor
\(\{\tilde{x}_{nj} = x_{nj}/M_n\}\) measuring proportions of total
mutation burden \(M_n\) attributable to individual variants \(\{j\}\)
and producing a normalized version \(\tilde{X}\) of the variant design
matrix. This normalization acknowledges mutation burden heterogeneity
observed in large-scale genomic data. Furthermore, it aids a product
matrix \(\tilde X U\) whose columns catalog analogous proportions of
\(\{M_n\}\) attributable to individual meta-feature categories. Next, we
embed into the hidden genome model a hierarchical topic model layer for
interpretable dimension reduction of the mutation context categories.
For notational simplicity below we restrict to a \emph{single,
categorical meta-feature source} (e.g., only SBS) for the model; a
generalization with several categorical and continuous meta-feature
sources is described in Section \ref{sec:multi-mfeat-sources}. The
proposed \emph{topical hidden genome model} is first succinctly
expressed as follows.

\begin{align}
\Pr(c_n &= k) = \frac{\exp\left[\alpha_k + e_{n;N}^T \tilde{X} \beta_{\bullet, k}\right]}{\exp\left[\sum_{k'=1}^K \alpha_{k'} + e_{n;N}^T\tilde{X}\beta_{\bullet, k'} \right]} \label{eq:model2:layer:mlogit} \\
\beta_{j, k} &= \beta^0_{j, k} + u_{j}^T \omega_{\bullet,k} \label{eq:model2:layer:mfeat-regr} \\
\omega_{\bullet, k} &= W^{-} \ \theta_{\bullet, k} \ \left(\implies W
\omega_{\bullet, k} = \theta_{\bullet, k}; \ W^{-} = W^T(WW^T)^{-1}\right) \label{eq:model2:layer:topic-regr} \\
\alpha_k &\sim \operatorname{N}(0, \tau_{0;\alpha}^2) \label{eq:model2:layer:icpt-prior} \\
\beta^0_{jk} &\sim \operatorname{N}\left(0,  \frac{\tau^2_{j;\beta^0}}{\tilde\sigma_{\text{obs}, j}^2} \right); \ \theta_{sk} \sim \operatorname{N}\left(0, \frac{\tau^2_{s;\theta}}{\tilde\sigma_{\text{topic}, s}^2} \right);
\tau^2_{j;\beta^0}, \tau^2_{s;\theta} \sim \operatorname{Gamma}\left(\frac{K+1}{2}, \frac{\lambda^2}{2}\right) \label{eq:model2:layer:gr-lasso}   \\
& \text{with }  \tilde\sigma_{\text{obs}, j} = \hat{\text{sd}}\left(\tilde X_{\bullet, j}\right); \  \tilde\sigma_{\text{topic}, s} = \hat{\text{sd}}\left(\left[\tilde X U W^{-}\right]_{\bullet, s}\right) \nonumber \\
\lambda^2 &\sim \operatorname{Gamma}(a_\lambda, b_\lambda) \label{eq:model2:layer:gr-lasso-lambda}  \\
(XU)_{np} &= \sum_{s=1}^S Z_{nsp}; \
Z_{nsp} \sim \operatorname{Poisson}\left( H_{ns} W_{sp}\right); \text{with } H_{ns}, W_{sp} \geq 0; \sum_{p=1}^P W_{sp} = 1  \label{eq:model2:layer:topic} \\
\Ht_{ns} &\sim \operatorname{Gamma}(a_H, b_H); \Wt_{sp} \sim \operatorname{Gamma}(a_W, b_W); \text{ then normalize } \label{eq:model2:layer:topic-prior} \\
W_{sp} &= \Wt_{sp}/\sum_{p'=1}^P W_{sp'}; \ H_{ns} = \Ht_{ns}\sum_{p'=1}^P W_{sp'}; \ \text{ so that }  H_{ns} W_{sp} = \Ht_{ns} \Wt_{sp} \label{eq:model2:layer:topic-normalize-W} \nonumber.
\end{align}

\noindent In above \(A_{\bullet, r}\) denotes the \(r\)-th column of a
matrix \(A\). Below we discuss the model in detail.

\hypertarget{sec:sig-hidgen-mlogit-layer}{%
\subsubsection{Multi-logit regression and generalized odds
ratios}\label{sec:sig-hidgen-mlogit-layer}}

The two topmost layers \eqref{eq:model2:layer:mlogit} and
\eqref{eq:model2:layer:mfeat-regr} above are similar to the original
hidden genome model. The model uses a symmetric multi-logistic
representation (Friedman, Hastie, and Tibshirani 2010) that allocates a
regression coefficient to each variant \(j\) and cancer site \(k\). The
model is \emph{not} likelihood identified and it requires the subsequent
layers for estimability of the regression parameters. Post-estimation
interpretation of the parameters can be made through a post-hoc
constraint, e.g., \(\sum_{k=1}^K\beta_{jk} = 0\) for each \(j\). Then
\(\beta_{jk}\) measures the ``one-vs-all'' log \emph{generalized} odds
ratio of classifying into the \(k\)-th cancer type relative to
\emph{all} \(K\) sites, given a one-unit change only in the
\emph{proportion of mutations} at variant \(j\) (Zahid and Tutz 2013).
Other types of generalized odds ratios, including ``one-vs-one'' (i.e.,
the usual baseline model odds ratios) and ``one-vs-rest'' odds ratios,
can also be derived from \(\{\beta_{jk}^0\}\), \(\{\theta_{sk}\}\), and
\(\{\omega_{pk}\}\); see Supplement A for more details.

\hypertarget{sec:sig-hidgen-mfeat-regr-layer}{%
\subsubsection{Observed and topic meta-feature
regression}\label{sec:sig-hidgen-mfeat-regr-layer}}

The layer \eqref{eq:model2:layer:topic-regr} expresses the
\emph{observed meta-feature} regression coefficients
\(\{\omega_{pk}: p=1, \dots, P\}\) as linear combinations of
\emph{topic-meta feature regression coefficients}
\(\{\theta_{sk}: s = 1, \dots, S\}\) corresponding to \(S\) latent
topics (\(S<P\); often \(S \ll P\); see the note in Supplement B). Here
\(W^{-} = W^T(WW^T)^{-1}\) so that \(WW^{-} = I_S\), the \(S\)
dimensional identity matrix. This construct induces a unique
\(\omega_{\bullet, k}\) satisfying
\(W \omega_{\bullet, k} = \theta_{\bullet, K}\) in that for any other
\(\tilde{\omega}_{\bullet, k}^*\) satisfying
\(W\tilde{\omega}_{\bullet, k}^* = \theta_{\bullet, k}\) we may recover
\(\omega_{\bullet, k}\) by projecting \(\tilde{\omega}_{\bullet, k}^*\)
on the row space of the topic matrix \(W\):
\(\omega_{\bullet, k} = W^T(WW^T)^{-1} W \tilde{\omega}^*_{\bullet, k} = \mathcal{P}_{W^T} \tilde{\omega}^*_{\bullet, k}\)
where \(\mathcal{P}_{W^T}\) denotes the projection matrix on the column
space of \(W^T\) or equivalently on the row space of \(W\). Furthermore,
since each row of \(W\) is a categorical probability measure embodying a
topic, \(\theta_{s, k}\) can be interpreted as \emph{the
expected/average} value of the observed meta-feature regression
coefficients \(\{\omega_{p, k}: p=1, \dots, P\}\) with respect to the
\(s\)-th topic \(\{W_{sp}: p = 1, \dots, P\}\). The connection
\eqref{eq:model2:layer:topic-regr} aids quantification of regression
coefficients associated with both observed meta-feature categories and
latent meta-feature topics in the model. This is in contrast to
virtually every existing supervised topic model in that they only aid
quantification for the latter; in genomic applications the former may
also be of relevance (see the example in Section \ref{sec:fit-infer}).
Note that \(\{\omega_{pk}\}\) can be derived post-hoc; for
implementation one focuses only on \(\{\theta_{sk}\}\) which aids
substantial dimension reduction over \(\{\omega_{pk}\}\).

\hypertarget{sec:sig-hidgen-regr-prior-layer}{%
\subsubsection{Hierarchical group-lasso prior on the independent
regression coefficients}\label{sec:sig-hidgen-regr-prior-layer}}

The hierarchical group lasso prior (Kyung et al. 2010) layer
\eqref{eq:model2:layer:gr-lasso} assigned on the \emph{independent}
regression coefficients \(\{\beta^0_{jk}, \theta_{sk}\}\) encourages
group-wise shrunken estimation of these parameters. Here each group is
constituted by the \(K\) coefficients across all cancer sites for a
variant \(j\) or a meta-feature topic \(s\). Because the corresponding
columns of \(\tilde X\) or \(\tilde X U W^{-}\) have different scales,
the group lasso prior is considered on coefficients \emph{scaled} by the
corresponding column sample standard deviations (denoted as
\(\hat{\text{sd}}(\cdot)\) and evaluated from the \(N\) data points).
Note that the sample standard deviations of the columns of
\(\tilde X U W^{-}\) involve model parameters \(W\), and hence this
scaling cannot be done simply as data pre-processing. In the proposed
MCMC sampler for the model this is done once per iteration; see Section
\ref{sec:implementation-mcmc} and supplement A for more details on
implementation of the model. To aid Bayesian estimation of the
group-lasso sparsity parameter \(\lambda\), a
\(\text{Gamma}(a_\lambda, b_\lambda)\) prior is considered for
\(\lambda^2\) on layer \eqref{eq:model2:layer:icpt-prior} (Kyung et al.
2010). When knowledge on the sparsity level is lacking a vague prior
induced, e.g., by \(a_\lambda = 0.01\) and \(b_\lambda = 0.01\) can be
used. Note that intercepts \(\{\alpha_k\}\) are not shrunken; they are
assigned a fixed, non-hierarchical n
\(\operatorname{N}(0, \tau_{0;\alpha}^2)\) prior in layer
\eqref{eq:model2:layer:icpt-prior}. A moderately large
\(\tau_{0, \alpha}\) (e.g., \(\tau_{0;\alpha} = 10)\) is suggested to
make the prior weakly informative.

\hypertarget{sec:sig-hidgen-mmsig-layer}{%
\subsubsection{Topic model for observed meta-feature
categories}\label{sec:sig-hidgen-mmsig-layer}}

The layer \eqref{eq:model2:layer:topic} induces a topic model over the
\(P\) observed meta-feature categories. To see this, consider tumor
\(n\in \{1, \dots, N\}\) with total mutation burden \(M_n\) of which
\(V_{np} = (XU)_{np}\) associates with the meta-feature category
\(p \in \{p = 1, \dots, P\}\). The quantities \(\{Z_{nsp}\}\) measure
\emph{latent contributions} of the \(s\)-th topic (\(s= 1, \dots, S\);
\(S < P\)) to the count \(V_{np}\) and aid conjugacy for MCMC sampling
(Supplement C; also see Liang and Hoffman (2014)). It follows that
\(V_{np} \sim \text{Poisson}(\sum_{s=1}^S H_{ns} W_{sp})\) and
conditional on \(M_n = \sum_{p=1}^P V_{np}\),
\((V_{n1}, \dots, V_{nP}) \sim \text{Multinomial}_P(M_n; (\sum_{s=1}^S \zeta_{ns} W_{s1}, \dots, \sum_{s=1}^S \zeta_{ns} W_{sP}))\)
where \(\zeta_{ns} = H_{ns}/\sum_{p=1}^P \sum_{s'=1}^S H_{ns'} W_{s'p}\)
\(= H_{ns}/\sum_{s'=1}^S H_{ns'}\). Equivalently,
\((V_{n1}, \dots, V_{nP}) \sim \sum_{s = 1}^S \zeta_{ns} f_s\) where
\(f_s \equiv \text{Multinomial}_P (M_n;\) \(( W_{s1}, \dots, W_{s1}))\).
Define topic-\(s\) through the \emph{composition probabilities}
\((W_{s1}, \dots, W_{s1})\). Then \(f_s\) is the probability
distribution of allocations assigning \(M_n\) mutations into \(P\)
meta-feature categories following the composition probabilities of
topic-\(s\), and \(\zeta_{ns}\) is understood as the \emph{exposure} to
topic-\(s\) in tumor \(n\).

We make a few notes here. First, \eqref{eq:model2:layer:topic} can be
interpreted as a parametric formulation of non-negative matrix
factorization (NMF) \(V \approx H W\) (Paisley, Blei, and Jordan 2014)
commonly used for mutation signature estimation in genomics. Second, the
parameters \(H\) and \(W\) are not identified. This makes subsequent
inference on them challenging. This is a well-known problem in topic
modeling and mixture models in general; in Section \ref{sec:infer-topic}
we discuss a simple clustering-based approach to this problem. Finally,
the topic model hyper-parameters, namely \(a_H\), \(b_H\); \(a_W\),
\(b_W\); and \(S\) all need to be judiciously chosen. A detailed
discussion on these parameters is provided in the Supplement B. In our
experiments we found small values such as \(a_H = b_H = 1\) and
\(a_W = b_W = 0.5\) and a moderately large \(S\) between \(50\) to
\(75\) to produce models balancing predictive ability, interpretability,
and computational costs.

\hypertarget{sec:implementation-mcmc}{%
\subsection{Full Bayesian Implementation via
MCMC}\label{sec:implementation-mcmc}}

The posterior distribution for the proposed model is highly intractable.
For implementation we propose a doubly data augmented, blocked,
\emph{independence} Metropolis-within-Gibbs algorithm for MCMC sampling
from the target posterior. We provide a summary of the algorithm below;
a detailed description with notes on initialization is provided in
Supplement A. We note that to aid computation, prior to using the MCMC
sampler one needs to screen the columns of the training predictor matrix
\(\tilde X\) \emph{after} computing the product matrices \(XU\) and
\(\tilde XU\). This effectively amounts to setting the residual effects
\(\{\beta_{jk}^0\}\) to be zero \emph{apriori} for all but a few (say
\(\leq 50\)) highly discriminating variants \(\{j\}\). Following
Chakraborty, Begg, and Shen (2020) we use a computationally efficient
mutual information based variable screening for this.

The proposed Gibbs-type MCMC algorithm iteratively draws (a) topic model
parameters, and (b) group-lasso multi-logistic parameters in two blocks
each containing a separate data-augmentation step. In block (a) latent
Poisson data \(\{Z_{nsp}\}\) (Liang and Hoffman 2014) are augmented to
aid MCMC sampling of \((H, W)\). To account for the \emph{supervised}
contribution of the multi-logistic layer we propose an
\emph{independence} Metropolis-Hastings step that \emph{requires no
manual tuning}. In block (b) a Gibbs draw is performed for the
hierarchical group-lasso (Kyung et al. 2010) multi-logistic parameters
with a \emph{column-scaled} \((\tilde{X}, \tilde{X}UW^{-})\) used as
predictor matrix (see Chakraborty, Begg, and Shen 2020). Conditional
conjugacy for the multi-logistic parameters are achieved via Pólya-Gamma
data augmentation (Polson, Scott, and Windle 2013).

\paragraph{Approximations}

While exact MCMC sampling along the steps above is theoretically
possible, it is still too computation-intensive to be practicable in
sizable applications. Fortunately, substantial computational gains are
achieved via a \emph{first order Taylor} approximation resembling the
usual NMF assumption
\(V_{np} \approx E(V_{np} \mid \{H_{ns}, W_{sp}\}) = \sum_{s=1}^S H_{ns}W_{sp}\)
(Lee and Seung 1999) inside the Metropolis-Hastings acceptance
probability for block (a). A second approximation to encourage better
mixing is made by assuming that the change in the standard deviation
terms \(\{\tilde{\sigma}_{\text{topic}, s}\}\) due to a change in \(H\)
(Metropolis Hastings proposal vs.~current values). Together, these
permit approximate, but efficient independent drawing of the elements of
\(W\) given \(H\) and the rows of \(H\) \emph{given} \(W\) (see
Supplement A).

\hypertarget{sec:multi-mfeat-sources}{%
\subsection{Handling multiple, possibly continuous, meta-feature
sources}\label{sec:multi-mfeat-sources}}

To handle multiple categorical meta-feature sources, one may introduce
into the model separate, independent topic model layers
\eqref{eq:model2:layer:topic}-\eqref{eq:model2:layer:topic-prior} and a
topic-to-observed coefficient connection layer
\eqref{eq:model2:layer:topic-regr} for each source. Continuous
meta-feature sources, if present, would directly produce observed
meta-feature regression coefficients \(\omega\). All \emph{observed}
meta-feature regression coefficients are then stacked for use in layer
\eqref{eq:model2:layer:mfeat-regr}. The MCMC algorithm discussed in
Section \ref{sec:implementation-mcmc} requires small modifications to
reflect these new layers. With several categorical meta-feature sources
one would sequentially update the topic parameters for each source
conditional on the ``supervised'' multi-logistic contributions of the
other source topics. The hierarchical group-lasso multi-logistic
regression is performed on an expanded column scaled predictor matrix
column-stacking \(X\), all \(\tilde{X}UW^{-}\) matrices arising out of
categorical meta-feature sources, and all \(XU\) matrices arising out of
continuous sources.

\hypertarget{sec:post-predict}{%
\subsection{Bayesian Posterior Predictive
Probabilities}\label{sec:post-predict}}

A full Bayesian ``ensemble'' prediction using posterior MCMC draws
\(\{\alpha^{(t)}, \beta^{0, (t)}, \omega^{(t)}: t = 1, \dots, T\}\) for
the multi-logistic parameters is described as follows. Given a
``new''/test tumor with \emph{normalized} variant indicators
\(\tilde x_\text{new}\), one first finds variants \(\{j\}\) in
\(\tilde x_\text{new}\) with non-zero residual effects
\(\{\beta_{j, k}^{0, (t)}: k=1, \dots, K\}\) for each draw
\(t=1, \dots, T\). All remaining variants \(\{j^*\}\), including
``novel'' variants in test data and screened out variants in training
data are assigned \(\{\beta_{j^*, k}^{0, (t)} = 0: k=1, \dots, K\}\).
The meta-feature values \(u_{j*}\) for a ``novel'' variant \(j^*\) are
cataloged as a new row of \(U\) for computation of
\(\tilde x_\text{new}^T U\). Subsequently, for each draw
\(t \in \{1, \dots, T\}\) one computes first the linear predictions
using the layers \eqref{eq:model2:layer:mlogit} and
\eqref{eq:model2:layer:mfeat-regr}:
\(\{\eta_{k}^{(t)} = \alpha_k^{(t)} + \tilde x_{\text{new}}^T\beta^{0, (t)} + \tilde x_{\text{new}}^T U \omega^{0, (t)} : k = 1, \dots, K \}\)
for all \(K\) cancer sites. These draw-specific linear predictions are
then ``softmax''-ed and subsequently averaged across draws to compute
cancer-site specific posterior predictive probabilities: \[
\begin{aligned}
&\quad \Pr(c_{\text{new}} = k \mid x_{\text{new}}, \text{training data}) \\
&= \int \Pr(c_{\text{new}} = k \mid \alpha, \beta^0, \omega,  \tilde x_{\text{new}}, \text{training data}) \ \pi(\alpha, \beta^0, \omega \mid \text{training data}) \ d(\alpha, \beta^0, \omega) \\
&\stackrel{\text{computed}}{=} \frac{1}{T} \sum_{t=1}^T \Pr(c_{\text{new}} = k \mid \alpha^{(t)}, \beta^{0, (t)}, \omega^{(t)}, x_{\text{new}}, \text{training data}) = \frac{1}{T} \sum_{t=1}^T \frac{\exp\left(\eta_{k}^{(t)}\right)}{1+ \exp\left(\eta_{k}^{(t)}\right) }.
\end{aligned}
\] Note that unlike a point estimate-based prediction (e.g., MAP
prediction) the above acknowledges model estimation uncertainty, and is
thus preferred in full Bayesian inference.

\hypertarget{sec:infer-topic}{%
\subsection{Inference on the latent signatures: post-hoc identification
via clustering}\label{sec:infer-topic}}

The topic related parameters \(W\), \(H\), and \(\{\theta_{sk}\}\) may
reveal important biological insights. However, their estimation is
challenging due to their non-identifiability caused by the mixture
structure of the topic model. Directly using MCMC outputs for these
parameters for point/interval estimation may produce misleading
results.\footnote{Note only direct inference on the (topic) parameters $H$, $W$, and $\{\theta_{sk}\}$ is challenging. Inference on posterior quantities that marginalizes over these parameters, e.g., prediction for new data and estimation of $\{\beta^0_{jk}\}$ and $\{\omega_{pk}\}$ is straightforward.}
Instead, using a point process representation of MCMC draws
(Frühwirth-Schnatter 2011), we introduce a simple clustering approach to
post-hoc identification and approximate inference for these parameters.

Briefly, we first create a list of all topics (the rows of \(W\)) drawn
across all iterations of the MCMC run, and cluster these pooled rows
using \(k\)-means with an appropriately chosen \(k\) (via elbow method
in our computations). The key idea is then to use the consequent cluster
assignments to relabel MCMC draws for the topics (rows of \(W\)), topic
exposures (columns of \(H\)) and topic regression coefficients
\(\{\theta_{sk}\}\). These cluster-relabeled draws are then treated as
posterior MCMC draws for ``post-hoc identified'' parameters and used for
MCMC output-analysis as usual. We make a few notes on the use of
\(k\)-means as a clustering algorithm in the current context. First, a
pairwise distance based clustering algorithm, e.g., \(k\)-medoids, might
be conceptually preferable to \(k\)-means given the simplex nature of
the topics. However they are not practicable when the number of topics
\(S\) and/or the number of MCMC iterations \(T\) is even moderately
large. Second, \(k\)-means results are sensitive to outliers. Here
outliers correspond to posterior draws for the rows of \(W\) that are
far from any other rows of \(W\) drawn in any iteration, and hence may
represent configurations with low posterior probabilities. To avoid the
influence of these outliers, we considered an approximate nearest
neighbors-based filtering prior to \(k\)-means clustering. Finally,
\(k\) means algorithm suffer from high computational expense and reduced
stability as the number of meta-feature categories (number of columns of
\(W\)) grows. This is particularly challenging for gene and window
meta-feature sources with hundreds to thousands of categories. To manage
the computation load, prior to \(k\)-means (and outlier detection) we
perform PCA and retain the first few (50 in our computations) principal
components.

The above approach resembles the \(k\)-means finite mixture model
identification approach of Frühwirth-Schnatter (2011). However, instead
of only resolving label-switching/permutation modes as done in finite
mixtures here we use clustering to also possibly combine duplicated
topics/modes arising \emph{within the same MCMC iteration}. Furthermore,
while heuristic, Bayesian estimation via ``cluster identified'' MCMC
draws may still be viewed as an approximate solution to a rigorously
defined decision theoretic problem; see Stephens (2000).

\hypertarget{sec:example}{%
\section{Data Example}\label{sec:example}}

We implement the proposed model on the full PCAWG whole genome data with
\(K = 10\) sites, and its restricted whole exome and targeted cancer
gene panel subsets. On the whole genome set we consider three discrete
meta-feature sources: gene, SBS, and 1-MB chromosome windows. For the
simulated subsets where chromosome windows trace only the gene level
mutations, we consider two sources: gene and SBS. For each source we set
\(S = 50\). The models are then fitted using the proposed MCMC algorithm
using marginal MAP (for \(\alpha\), \(\beta^0\), and \(\omega\)) and
unsupervised NMF (for \(H\) and \(W\)) based initialization of
parameters as described in Supplement D. On each dataset the MCMC
algorithm is run for 20,000 iterations after discarding the initial 1000
iterations as burn-in. For compute and memory feasibility, we updated
the topic model parameters \(H\) and \(W\) once every 10-th iterations
while updating all other parameters at every iteration; these 10-th
iterations are then stored \emph{(thinning)} and subsequently used for
Bayesian estimation and prediction.

\hypertarget{sec:cv-predict}{%
\subsection{Predictive performance under cross
validation}\label{sec:cv-predict}}

\begin{figure}[!htb]
    \centering
    \includegraphics[width=\textwidth]{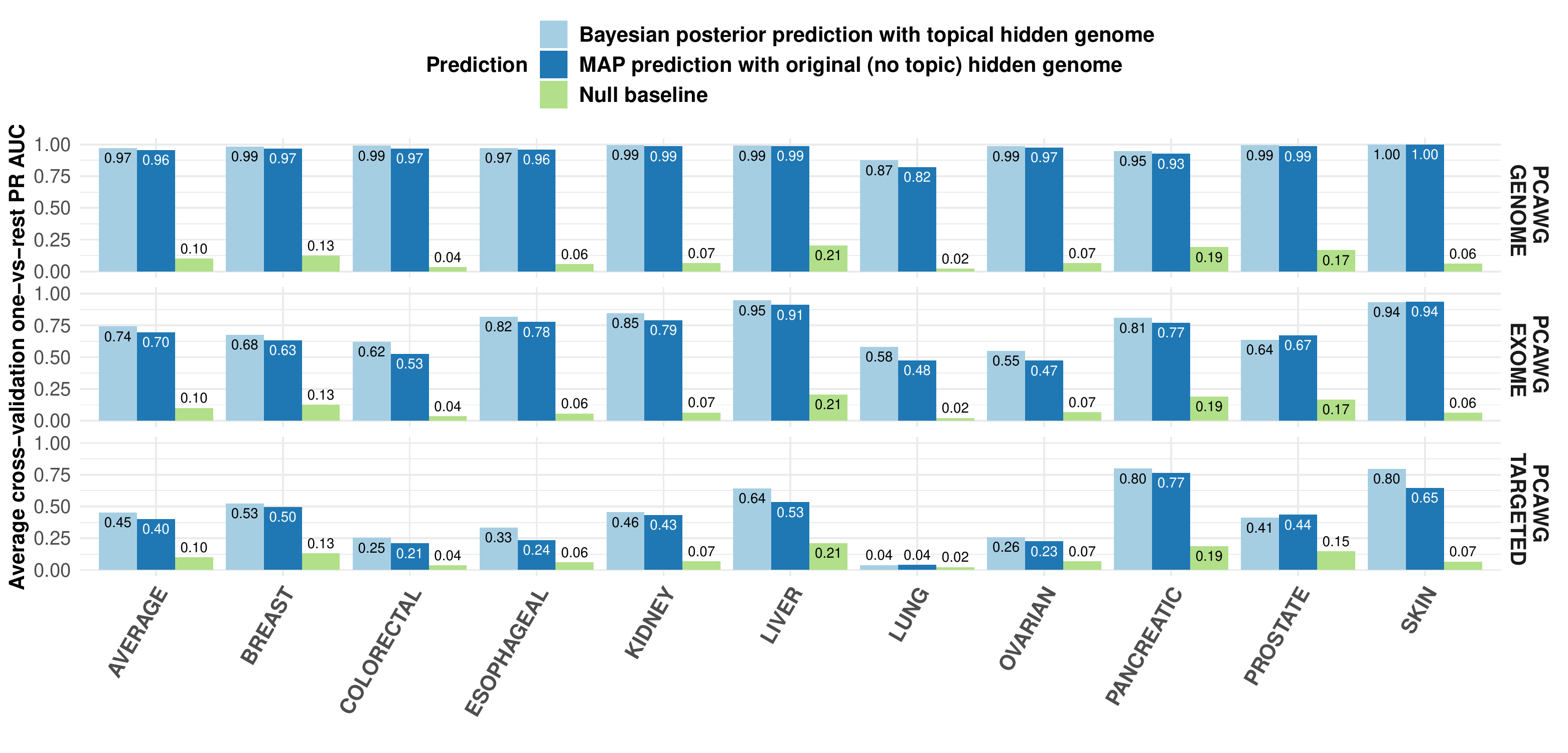}
    \caption{Comparing cross-validation PR AUC (averaged across 10 independent cross validations) of the original hidden genome model (without topics) MAP predictions and proposed topical hidden genome posterior predictive probabilities. The null baseline corresponds to a classifier that randomly assigns cancer site labels to tumors}
    \label{fig:pr_auc_compare}
\end{figure}

On each dataset, we perform 10 fold cross-validations to assess
predictive performance of the model. The folds are created using
stratified (based on cancer sites) random partitions of the PCAWG
tumors; these produce 10 different training/test set combinations
obtained by pooling each 9 of the 10 folds for training and the
remaining fold for testing. On each training set a separate model is
fitted and then used for Bayesian (posterior) prediction of the
corresponding test set. The predicted class probabilities for all tumors
in all folds are subsequently combined. To aid comparison, alongside we
also obtain analogous MAP predictions from the hidden genome model
(without any topics) of Chakraborty, Begg, and Shen (2020) with similar
normalized \(\tilde X\) as predictors. For each model, we derive
one-vs-rest classification probabilities for each of the \(K = 10\)
cancer sites, and compute the associated area under the one-vs-rest
precision-recall curve (PR AUC) as a measure of site-specific predictive
performance (Saito and Rehmsmeier 2015). PR AUCs reflect class-size
imbalances, and are more informative than ROC AUCs for one-vs-rest
comparisons obtained from a multi-class classifier. The site-specific PR
AUCs are then averaged to produce an overall metric for each model. We
consider 10 random replications of this cross-validation, and obtain
model specific average PR AUCs across replications. These average PR
AUCs are displayed as barplots in Figure \ref{fig:pr_auc_compare} with
results from the three different subsets plotted along panels. The
figure also shows the \emph{null baseline} PR AUC value for each site,
corresponding to a \emph{null} classifier randomly assigning
observations to classes. The null baseline PR AUCs are proportional to
the sample size of the corresponding class (see Chakraborty, Martin, et
al. (2021)).

We make the following observations from Figure \ref{fig:pr_auc_compare}.
First, the full Bayesian prediction from the proposed model has a better
overall-level performance than the MAP prediction from the original
model. The improvement is most prominent in targeted gene panel data
followed by exome level and whole genome level data. This overall
improvement is likely due to the topic model-based dimension reduction
which facilitates more stable estimation and prediction. Second, all
except one cancer site demonstrate improvement in site-specific
prediction in the proposed model. Certain cancer sites, such as lung and
pancreas, show consistently better prediction in all three datasets;
while for certain other sites, notably skin cancer in targeted gene
sequencing, the improvement is noticeable only at specific subsets.
Third, for only one site, namely prostate, is there a drop in PR AUC in
the proposed approach. This indicates the lack/non-existence of
discriminative topics for prostate cancer, which is known to have a
rather sporadic mutational landscape. However, the nominal drop in
prostate is more than compensated by improvements in the other sites,
making the proposed approach superior to the hidden genome model of
Chakraborty, Begg, and Shen (2020) as a predictive model.

\hypertarget{sec:fit-infer}{%
\subsection{Bayesian inference on tissue-specific topics and
meta-features and validation of results}\label{sec:fit-infer}}

This section illustrates how to make inference from the proposed model
though its applications on the PCAWG datasets. For brevity here we focus
on the full whole genome and the exome-subsetted datasets; analogous
inferences can also be drawn on the targeted cancer gene panel subsets.
Using the proposed MCMC algorithm we fit our model to each dataset
similarly as discussed in Section \ref{sec:cv-predict}, except now we
use the full datasets instead of cross-validation partitions to train
the models. Subsequently in each dataset we compute point (posterior
mean and median) and 80\% (highest posterior density) interval estimates
of topic model parameters and one-vs-rest log odds ratio parameters
obtained from regression coefficients for latent meta-feature topics,
observed meta-feature categories, and residual variant effects, using
the MCMC draws for model parameters.

\paragraph*{Tissue-specific window topics in PCAWG whole genome data. }

\begin{figure}[!htb]
    \centering
    \includegraphics[width=\textwidth]{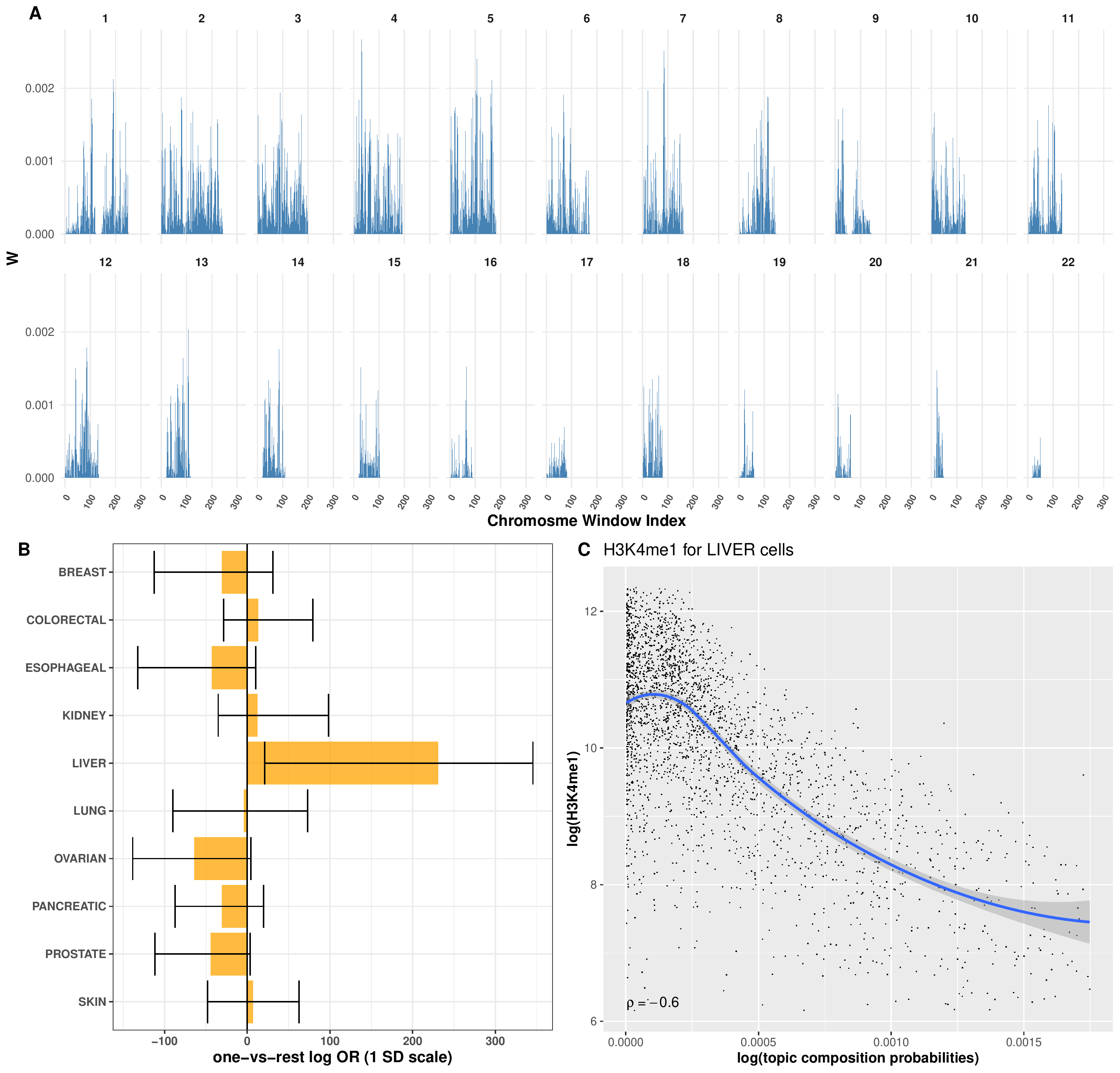}
    \caption{Inferring a liver cancer specific chromosome window topic estimated from the PCAWG data, and understanding its tissue site specificity and interpretability. \textbf{Panel A:} The blue vertical bars on  quantify posterior mean topic composition probabilities across $\sim 3000$ chromsome windows grouped by the chromosome numbers. \textbf{Panel B:} The orange horizontal bars and the black whiskered lines visualize posterior medians and 80\% HPD intervals for the cancer site-specific one-vs-rest log odds ratios,  calculated for one standard deviation increase in the total fraction of mutation burden attributable to the topic. \textbf{Panel C:} Points in the scatter display the log  composition probabilities for the topic along the x-axes and log epignomic features (histone marks H3k4me1) for liver cells along the y-axes for all $\sim 3000$ chromosome window meta-feature categories. Inscribed at the bottom-left corner of the panel is the Spearman correlation coefficient $\rho$ between  the corresponding topic composition probabilities and H3k4me1 scores.}
    \label{fig:window-topic-or}
\end{figure}

\quad We first focus on the whole genome dataset. It is known (Jiao et
al. 2020; Chakraborty, Martin, et al. 2021) that whole genome mutation
densities at 1-MB chromosome windows collectively aid near perfect
classification of tumor types, and so we obtain the estimated latent
window topics and their log odds ratios. Several highly tissue-specific
latent topics are obtained from the model with strikingly high log-odds
ratios. For exposition here we focus on one, a liver cancer specific
topic displayed on Figure \ref{fig:window-topic-or}. It is known that
somatic mutational landscape in the cancer genome is shaped by the
epigenome structure of the corresponding site of origin (Polak et al.
2015). For example, regions enriched for transcription regulatory
elements (e.g., histone marks) are observed to have low somatic mutation
rates. Indeed, a major pattern emerged from our analysis of the
whole-genome is that the variation of mutation density across the genome
can be explained to a high degree by site-specific epigenome
organization. Specifically, we compare the estimated composition
probabilities for this topic at each individual chromosome window with
an associated epigenomic histone score H3K4me1 obtained from cells
sequenced in an independent study cohort (Dunham et al. 2012) separately
for each of the 10 sites. Interestingly, these topic composition
probabilities have a moderately strong negative association with liver
cell-specific epigenomic features (Spearman \(\rho \approx -0.6\);
Figure \ref{fig:window-topic-or}C; a negative correlation indicates that
regions enriched for regulatory elements have low observed somatic
mutation density), but only small associations with other site-specific
epigenome features (Spearman \(\rho \leq 0.46\) in absolute values; see
Figure S1 in Supplement E). This implies that this topic may explain the
epigenomic landscape of liver tumors, and is thus biologically relevant
for characterizing liver cell of origin.

\paragraph*{Tissue-specific estimated SBS topics from PCAWG-exome data and assessing their reproducibility with estimates from TCGA-exome data and results from the literature.}

\begin{figure}[!htb]
    \centering
    \includegraphics[width=\textwidth,page=1]{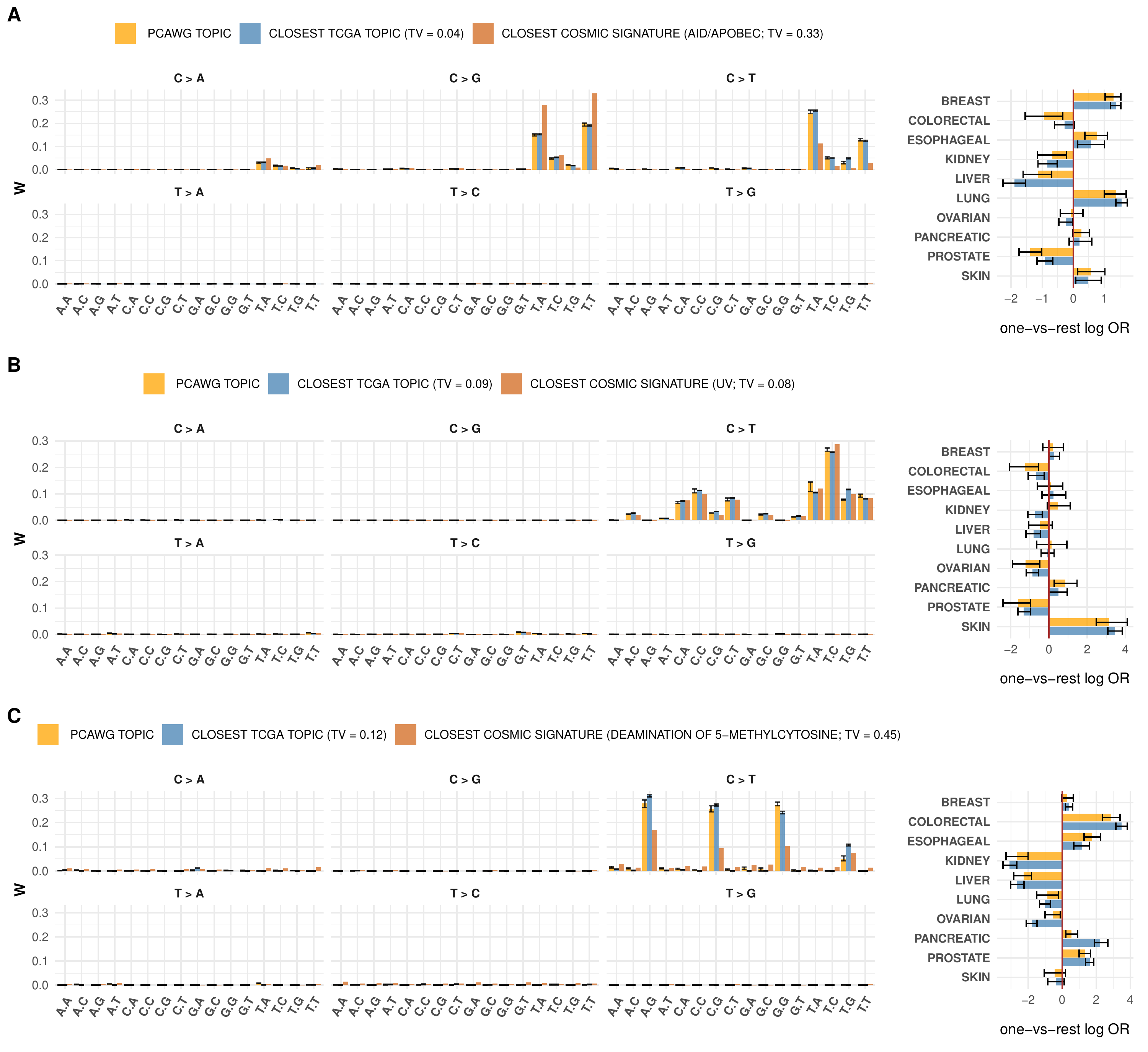}
    \caption{Inferring  3 strongly discrimative SBS topics (panels A-C) and their tissue site specificities estimated from the PCAWG data, and assessing reproducibility of the results with estimates from independent TCGA data and the COSMIC mutation signature repository. The vertical bars on the left show topic composition probabilitie, estimated via posterior means for PCAWG and TCGA datasets and collected from online repository for COSMIC, across 96 SBS categories grouped by subsitution types such as C$>$A, C$>$G etc. The horizontal bars on the right show estimated (via posterior median) cancer site specific one-vs-rest log odds ratios for one standard deviation increase in the total fraction of mutation burden attributable to the corresponding topic.  All PCAWG estimates are displayed as orange bars. For each PCAWG estimated topic, the closest (in terms of TV distance) TCGA  estimated (posterior mean) topic together with its estimated (posterior median) one-vs-rest log odds ratios are displayed through vertical and horizontal blue bars respectively, and the closest COSMIC mutation signatures are plotted as red vertical bars on the left. The names/functions of the closest COSMIC signatures are noted within parentheses in the legend. The black whiskers on each estimated bar display the associated 80\% HPD credible intervals. }
    \label{fig:sbs-topic-or}
\end{figure}

\quad We next focus on the model trained on the PCAWG exome subsetted
data, and obtain estimates for the SBS topics and the corresponding
tissue-specificities as quantified by the odds ratios. As noted before,
much research has been done on SBS signatures (analogous to topics) in
exome level mutation data, and several highly tissue site specific SBS
signatures have been detected. This permits direct reproducibility
assessment for our results. We consider (a) estimated SBS topics and
odds ratios from a similar model trained on the independent TCGA exome
data restricted to the same 10 cancer sites, and (b) exome level known
SBS mutation signatures noted on the literature (COSMIC database; Ludmil
B. Alexandrov et al.
(2020))\footnote{The mutation signatures noted in COSMIC are derived from the TCGA data using an unsupervised NMF approach, and they collectively serve as a ``baseline'' for comparing our PCAWG-estimated SBS topics.}
as a biological benchmark. We highlight top three discriminative
PCAWG-exome estimated SBS topics with high one-vs-rest log odds ratios,
and compare them with all TCGA-estimated SBS topics, and COSMIC
signatures to identify the closest match for each. Here ``closeness'' is
measured through total variation distance defined for two topic
compositions (discrete probability functions) \(W_1\) and \(W_2\) over
the same meta-feature categories \(\{1, \dots, P\}\) as
\(\text{TV}(W_1, W_2) = \frac{1}{2} \sum_{p=1}^P |W_{1p} - W_{2p}|\).
These three topics are displayed on the three panels of Figure
\ref{fig:sbs-topic-or}. The second topic (panel B) has a very close
match to the UV signature in the COSMIC database and uniquely associated
with melanoma (skin). The first topic (panel A) show a pattern that
clearly matches the APOBEC signature from COSMIC, but the TV distance is
relatively higher suggesting this topic is a modified version of APOEBEC
signature. Since APOBEC exist in multiple cancer types, this difference
is likely attributable to our supervised topic model contrasting a
cancer-site agnostic approach used to derive the COSMIC signatures.
Similarly, the third topic (panel C) shows a pattern that matches the
COSMIC signature of De-amination of 5-methylcytosine (clock-like), but
have relatively poorer matches in TV distance since we incorporated
cancer site information in our analysis.

\paragraph*{Tissue-specific gene topics from PCAWG-exome subsetted data. }

\begin{figure}[!htb]
    \centering
    \includegraphics[width=1.05\textwidth]{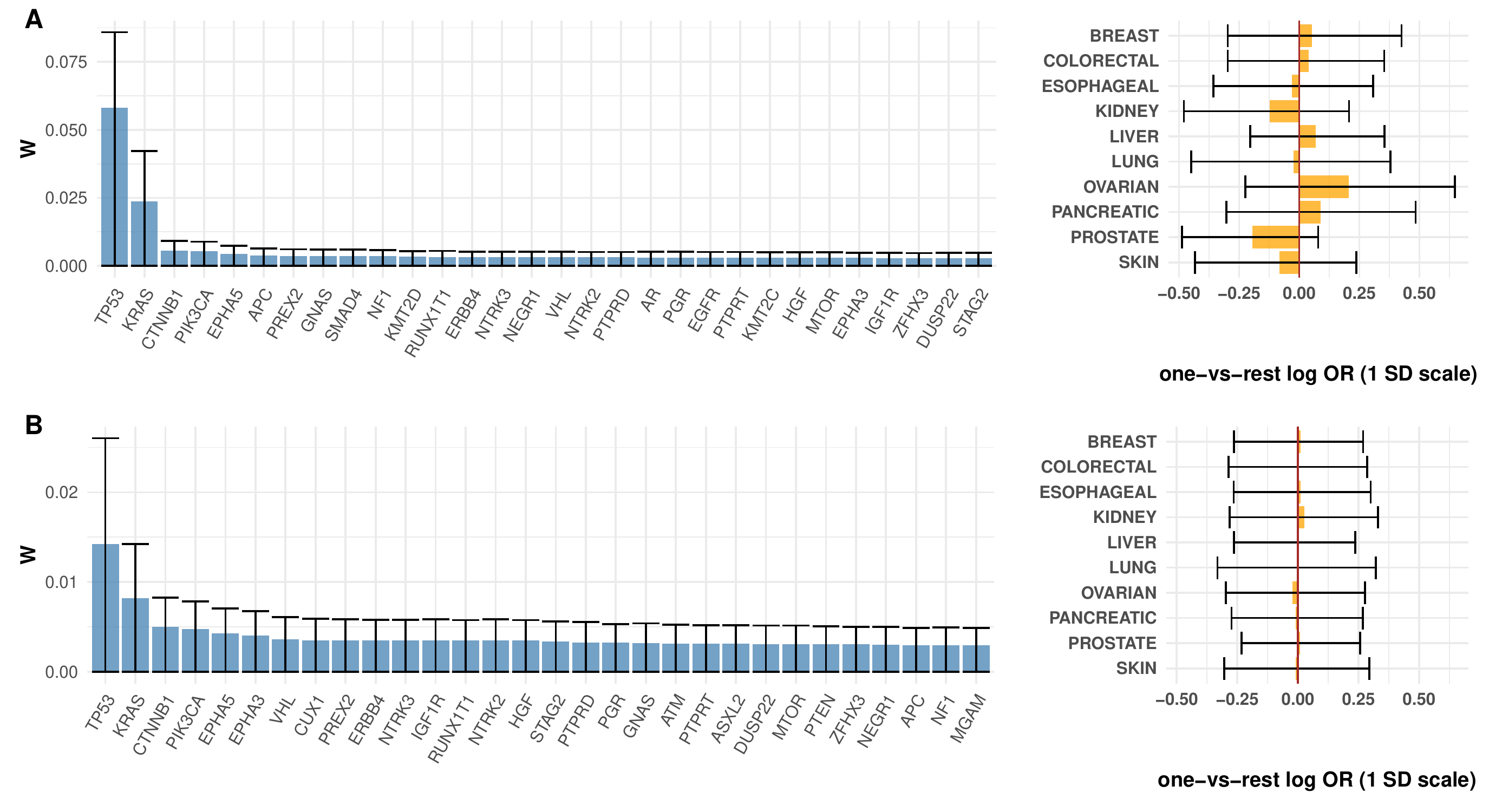}
    \caption{Inferring  2 gene topics and their tissue site specificities for PCAWG exome subsetted data. The blue vertical bars on the left quantify posterior mean topic composition probabilities for 30 most prominent genes in the topic, and the black vertical error-whisker bars show associated 80\% HPD intervals for the weights. The orange horizontal bars and the black whiskered lines on the right visualize posterior medians and 80\% HPD intervals for the cancer site-specific one-vs-rest generalized odds ratios,  calculated for one standard deviation increase in the total fraction of mutation burden attributable to the corresponding topic.}
    \label{fig:gene-topic-or}
\end{figure}

\quad Interestingly, the gene meta-feature categories did not produce
strong tissue specific latent topics, and the estimated topics showed
high variability in their composition probabilities. In Figure
\ref{fig:gene-topic-or} we display two gene topics with the largest
median site specific one-vs-rest log ORs. Both topics are strongly
driven by TP53 and KRAS, although the composition probabilities vary
substantially (vertical error bar over the bars). Furthermore, for all
cancer sites, the 80\% CIs for the log ORs (horizontal whiskers on the
right panels) span almost entirety of (-0.5, 0.5), indicating the lack
of tissue-site specificity of these topics. Together these likely
indicate that the tissue-specific genes do not mutate in unison, unlike
the chromosome windows and SBS categories.

\paragraph*{Inferring tissue specificities of residual variants and "observed" meta-features. }

\begin{figure}[!htb]
    \centering
    \includegraphics[width=1.05\textwidth]{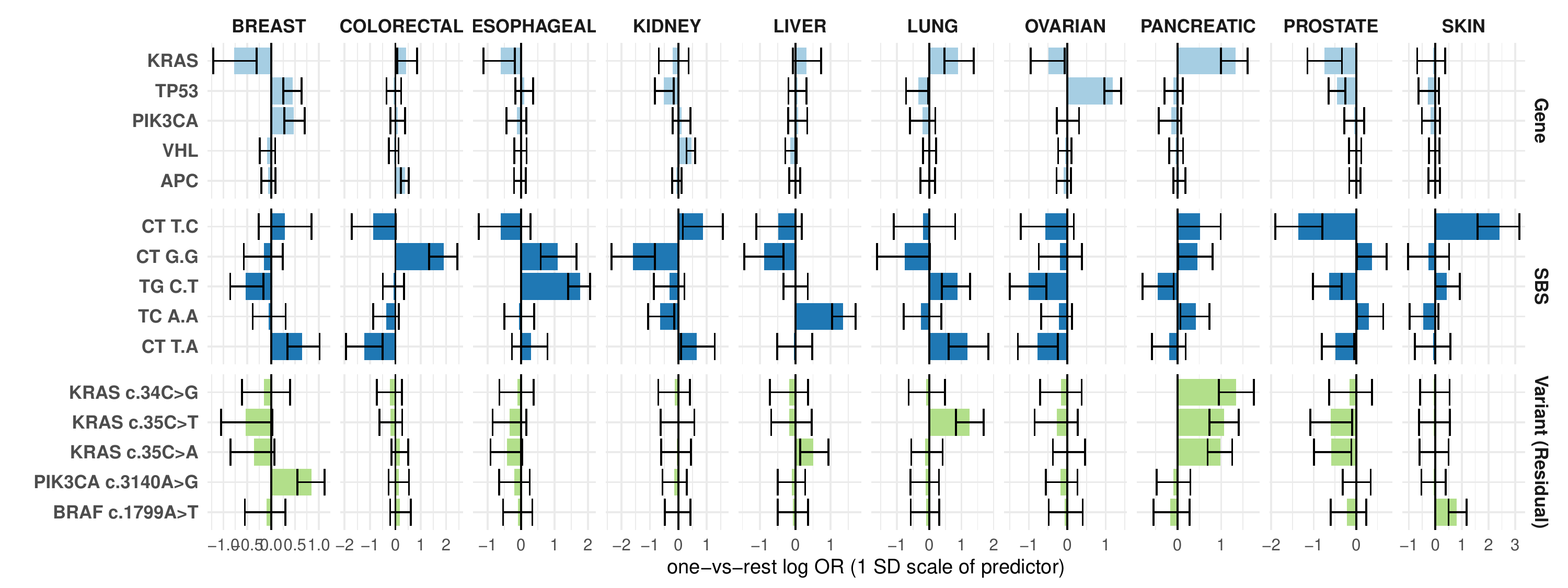}
    \caption{Inferring tissue site specificities attibutable to observed meta-feature categories, and residual variant effects. The horizontal bars and the black whiskered lines visualize posterior medians and 80\% HPD intervals for the cancer site-specific one-vs-rest generalized odds ratios,  calculated for one standard deviation increase in the total fraction of mutation burden attributable to each meta-feature category or each individual variant.}
    \label{fig:obs-pred-or}
\end{figure}

\quad Finally, we focus on individual variant-specific \emph{residual}
effects and ``observed'' meta-feature category effects. We consider the
exome level PCAWG data, and obtain component wise one-vs-rest log odds
ratios from the residual variant effects \(\beta^0\) and observed
meta-feature category effects \(\omega\). Then we rank the individual
variants and observed meta-feature categories (separately for each
meta-feature source of gene and SBS) based on their maximum (posterior
median) log one-vs-rest odds ratio values obtained across cancer sites.
The top five predictors from each group (variants and each meta-feature
source) are then selected, and the posterior median and 80\% HPD
intervals for their cancer site specific log one-vs-rest odds ratios are
plotted as horizontal bars in Figure \ref{fig:obs-pred-or}. The findings
visualized in Figure \ref{fig:obs-pred-or} agree with existing
scientific knowledge, and aid quantification of some interesting genomic
facts. For example KRAS gene mutations is known to be associated with
multiple cancer sites including pancreatic and lung; however, specific
variants of KRAS, e.g., c.35C\textgreater A are more specific to only
pancreatic cancer.

\hypertarget{sec:discuss}{%
\section{Discussion}\label{sec:discuss}}

This paper proposes a multilevel Bayesian multi-logit regression model
for statistical inference in ultra-high dimensional genome driven cancer
type characterization problems. The approach leverages bi-level
dimension reduction for efficient, practicable MCMC-based implementation
-- (a) condensing information in millions of observed, sparse variants
into a few thousands of observed meta-feature contexts, and then (b)
modeling the effects of these observed meta-features through lower
dimensional latent \emph{meta-feature topics}. The latter is a
manifestation of topic models (Blei and Lafferty 2009) that derive
\emph{interpretable} latent lower dimensional structures (``topics'')
from observed high-dimensional categorical data to facilitate
\emph{parts-based representation} of a complex underlying system. These
models have found important applications in several scientific and
engineering disciplines, including computational linguistics, genomics
(particularly, mutation signature extraction; Ludmil B. Alexandrov et
al. (2013); Funnell et al. (2019)), computer vision, and acoustic signal
processing, among others. The usefulness of topic models in aiding
interpretable lower dimensional structures has prompted its use as an
interpretable dimension reduction tool for regression/classification
(supervised learning) tasks, and a rich literature has been devoted to
the development of supervised topic models. See Mcauliffe and Blei
(2007), Zhu, Zheng, and Zhang (2013), Magnusson, Jonsson, and Villani
(2020) and the references therein for a review. To our knowledge,
however, none of the existing approaches integrate a topic model as an
intermediate layer of an elaborate Bayesian multinomial-logistic
regression model, and enable MCMC-based estimation, as proposed herein.
Furthermore, our approach permits quantification of regression effects
for both the observed meta-feature categories, and the latent
meta-feature topics. To the best of our knowledge this is not provided
by any existing supervised topic model approach; existing approaches
quantify regression coefficients of the latent topics only.

The topic model layer however introduces non-identified parameters in
the model, which subsequently makes statistical inference on these
parameters challenging. Herein we propose a \(k\)-means based posthoc
identification of the model, leveraging a point-process representation
of the MCMC draws as suggested in Frühwirth-Schnatter (2011), and our
results suggest reasonable practical performance of the proposed
intuitive approach.

There are several future directions we aim to pursue in both
methodological and applied scientific fronts. First, on the methodology
side, we aim to rigorize the proposed clustering-based posthoc
identification step for principled inference on the topic model
parameters. Second, while not considered herein, interest may lie in
allowing `information sharing' between cancer sites through
appropriately articulated hierarchical model layers. This may permit
more coherent estimation of the tissue-specific effects of mutations.
Third, a future research is planned to assess statistical performance of
the model via large scale simulations. Finally, on the scientific side,
we aim a thorough investigation of the tissue-specific whole genome
chromosome region topics to understand their biological relevance.

\hypertarget{references}{%
\section*{References}\label{references}}
\addcontentsline{toc}{section}{References}

\hypertarget{refs}{}
\begin{CSLReferences}{1}{0}
\leavevmode\vadjust pre{\hypertarget{ref-alexandrov2020}{}}%
Alexandrov, Ludmil B., Jaegil Kim, Nicholas J. Haradhvala, et al. 2020.
{``The Repertoire of Mutational Signatures in Human Cancer.''}
\emph{Nature} 578 (7793): 94101.

\leavevmode\vadjust pre{\hypertarget{ref-alexandrov2013}{}}%
Alexandrov, Ludmil B, Serena Nik-Zainal, David C Wedge, et al. 2013.
{``Signatures of Mutational Processes in Human Cancer.''} \emph{Nature}
500 (7463): 415--21.

\leavevmode\vadjust pre{\hypertarget{ref-bailey2018}{}}%
Bailey, Matthew H, Collin Tokheim, Eduard Porta-Pardo, Sohini Sengupta,
Denis Bertrand, Amila Weerasinghe, et al. 2018. {``Comprehensive
Characterization of Cancer Driver Genes and Mutations.''} \emph{Cell}.

\leavevmode\vadjust pre{\hypertarget{ref-blei2009}{}}%
Blei, David M., and John D. Lafferty. 2009. {``Topic Models.''} In.
Chapman; Hall/CRC.

\leavevmode\vadjust pre{\hypertarget{ref-campbell2020}{}}%
Campbell, Peter J., Gad Getz, Jan O. Korbel, Joshua M. Stuart, Jennifer
L. Jennings, Lincoln D. Stein, et al. 2020. {``Pan-Cancer Analysis of
Whole Genomes.''} \emph{Nature} 578 (7793): 82--93.

\leavevmode\vadjust pre{\hypertarget{ref-chakraborty2019}{}}%
Chakraborty, Saptarshi, Arshi Arora, Colin B. Begg, and Ronglai Shen.
2019. {``Using Somatic Variant Richness to Mine Signals from Rare
Variants in the Cancer Genome.''} \emph{Nature Communications} 10 (1).

\leavevmode\vadjust pre{\hypertarget{ref-chakraborty2020}{}}%
Chakraborty, Saptarshi, Colin B. Begg, and Ronglai Shen. 2020. {``Using
the {"}Hidden{"} genome to improve classification of cancer types.''}
\emph{Biometrics}, September.

\leavevmode\vadjust pre{\hypertarget{ref-chakraborty2021a}{}}%
Chakraborty, Saptarshi, Brett L. Ecker, Ken Seier, Victoria G. Aveson,
et al. 2021. {``Genome-Derived Classification Signature for Ampullary
Adenocarcinoma to Improve Clinical Cancer Care.''} \emph{Clinical Cancer
Research}, August.

\leavevmode\vadjust pre{\hypertarget{ref-chakraborty2021}{}}%
Chakraborty, Saptarshi, Axel Martin, Zoe Guan, Colin B. Begg, and
Ronglai Shen. 2021. {``Mining Mutation Contexts Across the Cancer Genome
to Map Tumor Site of Origin.''} \emph{Nat. Commun} 12 (1): 3051.

\leavevmode\vadjust pre{\hypertarget{ref-chakravarty2017}{}}%
Chakravarty, Debyani, Jianjiong Gao, Sarah Phillips, et al. 2017.
{``{OncoKB}: A Precision Oncology Knowledge Base.''} \emph{JCO Precision
Oncology}, no. 1 (November): 1--16.

\leavevmode\vadjust pre{\hypertarget{ref-dunham2012}{}}%
Dunham, Ian, Anshul Kundaje, Shelley F. Aldred, Patrick J. Collins,
Carrie A. Davis, Francis Doyle, et al. 2012. {``An Integrated
Encyclopedia of {DNA} Elements in the Human Genome.''} \emph{Nature} 489
(7414): 57--74. \url{https://doi.org/10.1038/nature11247}.

\leavevmode\vadjust pre{\hypertarget{ref-efron2020}{}}%
Efron, Bradley. 2020. {``Prediction, Estimation, and Attribution.''}
\emph{International Statistical Review} 88: S28S59.

\leavevmode\vadjust pre{\hypertarget{ref-friedman2010}{}}%
Friedman, Jerome H., Trevor Hastie, and Rob Tibshirani. 2010.
{``Regularization Paths for Generalized Linear Models via Coordinate
Descent.''} \emph{Journal of Statistical Software} 33 (February): 1--22.

\leavevmode\vadjust pre{\hypertarget{ref-fruhwirth2011label}{}}%
Frühwirth-Schnatter, Sylvia. 2011. {``Dealing with Label Switching Under
Model Uncertainty.''} In, edited by Kerrie L. Mengersen, Christian
Robert, and Mike Titterington, 213239. John Wiley \& Sons; Citeseer.

\leavevmode\vadjust pre{\hypertarget{ref-funnell2019}{}}%
Funnell, Tyler, Allen W. Zhang, Diljot Grewal, et al. 2019.
{``Integrated Structural Variation and Point Mutation Signatures in
Cancer Genomes Using Correlated Topic Models.''} \emph{PLOS
Computational Biology} 15 (2): e1006799.

\leavevmode\vadjust pre{\hypertarget{ref-haigis2019}{}}%
Haigis, Kevin M., Karen Cichowski, and Stephen J. Elledge. 2019.
{``Tissue-specificity in cancer: The rule, not the exception.''}
\emph{Science (New York, N.Y.)} 363 (6432): 1150--51.

\leavevmode\vadjust pre{\hypertarget{ref-jiao2020}{}}%
Jiao, Wei, Gurnit Atwal, Paz Polak, et al. 2020. {``A Deep Learning
System Accurately Classifies Primary and Metastatic Cancers Using
Passenger Mutation Patterns.''} \emph{Nat. Commun} 11 (1).

\leavevmode\vadjust pre{\hypertarget{ref-kyung2010}{}}%
Kyung, Minjung, Jeff Gill, Malay Ghosh, and George Casella. 2010.
{``Penalized Regression, Standard Errors, and Bayesian Lassos.''}
\emph{Bayesian Analysis} 5 (2): 369411.

\leavevmode\vadjust pre{\hypertarget{ref-lapuschkin2019}{}}%
Lapuschkin, Sebastian, Stephan Wäldchen, Alexander Binder, Grégoire
Montavon, Wojciech Samek, and Klaus-Robert Müller. 2019. {``Unmasking
Clever Hans Predictors and Assessing What Machines Really Learn.''}
\emph{Nature Communications} 10 (1): 1096.
\url{https://doi.org/10.1038/s41467-019-08987-4}.

\leavevmode\vadjust pre{\hypertarget{ref-lee1999}{}}%
Lee, Daniel D., and H. Sebastian Seung. 1999. {``Learning the Parts of
Objects by Non-Negative Matrix Factorization.''} \emph{Nature} 401
(6755): 788--91.

\leavevmode\vadjust pre{\hypertarget{ref-liang2014}{}}%
Liang, Dawen, and Matthew D. Hoffman. 2014. {``Beta Process Non-Negative
Matrix Factorization with Stochastic Structured Mean-Field Variational
Inference.''} \url{http://arxiv.org/abs/1411.1804}.

\leavevmode\vadjust pre{\hypertarget{ref-magnusson2020}{}}%
Magnusson, Måns, Leif Jonsson, and Mattias Villani. 2020. {``DOLDA: A
Regularized Supervised Topic Model for High-Dimensional Multi-Class
Regression.''} \emph{Computational Statistics} 35 (1): 175--201.
\url{https://doi.org/10.1007/s00180-019-00891-1}.

\leavevmode\vadjust pre{\hypertarget{ref-mcauliffe2007}{}}%
Mcauliffe, Jon, and David Blei. 2007. {``Supervised Topic Models.''} In.
Vol. 20. Curran Associates.
\url{https://proceedings.neurips.cc/paper/2007/hash/d56b9fc4b0f1be8871f5e1c40c0067e7-Abstract.html}.

\leavevmode\vadjust pre{\hypertarget{ref-paisley2014}{}}%
Paisley, John, David M. Blei, and Michael I. Jordan. 2014. {``Bayesian
Nonnegative Matrix Factorization with Stochastic Variational
Inference.''} In. Chapman; Hall/CRC.

\leavevmode\vadjust pre{\hypertarget{ref-polak2015}{}}%
Polak, Paz, Rosa Karlić, Amnon Koren, Robert Thurman, Richard Sandstrom,
Michael Lawrence, et al. 2015. {``Cell-of-Origin Chromatin Organization
Shapes the Mutational Landscape of Cancer.''} \emph{Nature} 518 (7539):
360--64. \url{https://doi.org/10.1038/nature14221}.

\leavevmode\vadjust pre{\hypertarget{ref-polson2013}{}}%
Polson, Nicholas G., James G. Scott, and Jesse Windle. 2013. {``Bayesian
Inference for Logistic Models Using {P}ólya{\textendash}Gamma Latent
Variables.''} \emph{J Am Stat Assoc} 108 (504): 1339--49.
\url{https://doi.org/10.1080/01621459.2013.829001}.

\leavevmode\vadjust pre{\hypertarget{ref-saito2015}{}}%
Saito, Takaya, and Marc Rehmsmeier. 2015. {``The Precision-Recall Plot
Is More Informative Than the {ROC} Plot When Evaluating Binary
Classifiers on Imbalanced Datasets.''} \emph{PLOS ONE}, March.

\leavevmode\vadjust pre{\hypertarget{ref-soh2017}{}}%
Soh, Kee Pang, Ewa Szczurek, Thomas Sakoparnig, and Niko Beerenwinkel.
2017. {``Predicting Cancer Type from Tumour DNA Signatures.''}
\emph{Genome Medicine} 9 (1).

\leavevmode\vadjust pre{\hypertarget{ref-stephens2000}{}}%
Stephens, Matthew. 2000. {``Dealing with Label Switching in Mixture
Models.''} \emph{Journal of the Royal Statistical Society: Series B
(Statistical Methodology)} 62 (4): 795--809.
\url{https://doi.org/10.1111/1467-9868.00265}.

\leavevmode\vadjust pre{\hypertarget{ref-yuan2006}{}}%
Yuan, Ming, and Yi Lin. 2006. {``Model Selection and Estimation in
Regression with Grouped Variables.''} \emph{J R Stat Soc Series B Stat
Methodol} 68 (1): 49--67.

\leavevmode\vadjust pre{\hypertarget{ref-zahid2013}{}}%
Zahid, Faisal Maqbool, and Gerhard Tutz. 2013. {``Ridge Estimation for
Multinomial Logit Models with Symmetric Side Constraints.''}
\emph{Computational Statistics} 28 (3): 1017--34.
\url{https://doi.org/10.1007/s00180-012-0341-1}.

\leavevmode\vadjust pre{\hypertarget{ref-zehir2017}{}}%
Zehir, Ahmet, Ryma Benayed, Ronak H Shah, Aijazuddin Syed, Sumit Middha,
Hyunjae R Kim, et al. 2017. {``Mutational Landscape of Metastatic Cancer
Revealed from Prospective Clinical Sequencing of 10,000 Patients.''}
\emph{Nature Medicine}.

\leavevmode\vadjust pre{\hypertarget{ref-zhu2013}{}}%
Zhu, Jun, Xun Zheng, and Bo Zhang. 2013. {``Improved Bayesian Logistic
Supervised Topic Models with Data Augmentation.''} \emph{arXiv Preprint
arXiv:1310.2408}.

\end{CSLReferences}

\end{document}


\maketitle

\allowdisplaybreaks

\hypertarget{sec:odds-ratio-defn}{%
\section{Different types of generalized odds ratios with generalized
baseline categories derived from the symmetric multinomial logistic
model}\label{sec:odds-ratio-defn}}

The generalized odds for a cancer site \(k\) given the baseline of
\emph{all cancer sites} \(\{1, \dots, K\}\) is defined as \[
\text{odds}(\{k\}\mid \{1, \dots, K\}) = \left. \Pr(c = k) \right/ \left\{\prod_{k'=1}^K \Pr(c = k')\right\}^{1/K}.
\] More generally, for a class \(A \subseteq \{1, \dots, K\}\) of cancer
sites of primary interest and a set \(B \subseteq \{1, \dots, K\}\) of
baseline cancer sites , we may define the generalized joint odds of the
set \(A\) relative to the baseline set \(B\) as
\[\text{odds}(A \mid B) = \left. \left\{\prod_{k\in A} \Pr(c = k)\right\}^{1/\#(A)} \right/ \left\{\prod_{k \in B} \Pr(c = k')\right\}^{1/\#(B)} = \left\{\prod_{k\in A} \text{odds}(\{k\} \mid B)\right\}^{1/\#(A)}\]
where \(\#(A)\) denotes the cardinality of \(A\). Then \[
\frac{1}{\#(A)} \sum_{k \in A} \beta_{jk} - \frac{1}{\#(B)} \sum_{k' \in B} \beta_{jk'}
\] would measure the log-ratio of changes in \(\text{odds}(A \mid B)\)
for an analogous one unit change in the predictor value of variant
\(j\). Note that by choosing \(B\) to be the set of one specific cancer
site say ``1'' we obtain joint odds ratios relative to cancer site
``1''. On the other hand for each cancer site \(k\), choosing
\(A = \{k\}\) and \(B = \{1, \dots, K\} \setminus \{k\}\) provides the
one-vs-rest generalized odds for site \(k\).

Analogous generalized odds ratio-based interpretations are obtained for
an observed meta-feature category \(p\) through coefficients
\(\{\omega_{pk}\}\), and a meta-feature topic \(s\) through coefficients
\(\{\theta_{sk}\}\); the odds ratios are here measured for one unit
change in the proportion of total mutation burden attributable to the
corresponding (topical) meta-feature category.

\hypertarget{sec:sig-hidgen-mmsig-prior-layer}{%
\section{The roles of hyperparameters in the topic model
priors}\label{sec:sig-hidgen-mmsig-prior-layer}}

Priors for the topic model parameters play important roles in
determining the distinctness and specificity of the produced topics in
individual tumors. At the outset we note that we work with the
non-normalized versions \(\tilde{H}\) and \(\tilde{W}\) of the topic
model parameters that satisfy \(\tilde{W}_{sp} \in (0, \infty)\) with
\(H_{ns} W_{sp} = \tilde{H}_{ns} \tilde{W}_{sp}\) and assign gamma
priors on them for analytical tractability in the proposed MCMC sampler.
However, our interest lies in the normalized versions \(\{H_{ns} \}\) or
more specifically \(\{\zeta_{ns}\}\) (tumor specific topic exposures)
and \[\{W_{ sp} = \tilde{W}_{ sp}/ \sum_{p=1}^P \tilde{W}_{ sp} \}\]
(topic specific meta-feature category weights); the prior rate
parameters \(b_H\) and \(b_W\) do not have any apriori impact on these
quantities and we hence may use a simpler choice, e.g.~\(a_H = b_H\) and
\(a_W = b_W\). Note that these gamma priors induce
\(\text{Dirichlet}_S(a_H 1_S)\) and \(\text{Dirichlet}_P(a_W 1_P)\)
priors on \(\{\zeta_{ns}\}\) and \(\{W_{sp}\}\) respectively where
\(\text{Dirichlet}_R\) denotes an \(R\) component Dirichlet distribution
and \(1_R\) is the \(R\) component vector of all ones.

Leveraging the consequent analogy with the Latent Dirichlet Allocation
(LDA) mixed-membership model (Blei, Ng, and Jordan 2003), it follows
that a smaller (larger) value of \(a_H > 0\) effectuates smaller
(larger) mixing weights for most topics within a tumor, making each
tumor exhibit fewer (more) topics with non-trivial exposures. On the
other hand a smaller (larger) value of \(b_H > 0\) makes each topic
assign smaller (larger) weights to most meta-feature categories, thus
making each individual topic more (less) specific to particular
meta-feature categories, and thus more (less) distinct. Because our
interest lies in obtaining meta-feature topics that are as distinct
(uncorrelated) and as tissue site specific as possible for interpretable
inference, we prefer smaller values of \(a_H\) and \(b_H\). However, too
small values of \(a_H\) and \(b_H\) may produce topics that are entirely
specific to individual tumors and are driven by only a few individual
meta-feature categories. This in turn may result in poor predictive
ability of the model and/or require a large number \(S\) of topics to
capture all tissue-specific information in all observed meta-feature
categories -- a situation that curtails the computational gain achieved
through the dimension reduction by signaturization. In theory, one may
impose some priors on \(a_W\) and \(b_W\) to let them estimate from the
data subject to some cost (loss) function balancing interpretability and
predictive ability and number of topics. However, this will require
additional computation-heavy steps in the proposed Gibbs sampler, and
care must be taken while defining an appropriate cost function. We leave
this direction as a potential future work and instead consider a
reasonable fixed \(a_H, b_H\) setting in the current paper. In all our
test applications across datasets the choice \(a_H = b_H = 1\) and
\(a_W = b_W = 0.5\) provided a reasonable balance between
interpretability and predictive ability of the model, and are thus used
as fixed prior in this paper.

The second hyper-parameter in the topic model that requires
pre-specification is the number of topics \(S\). To this, as rule of
thumb, we suggest setting \(S\) to some moderately large number such as
an integer between \(50\) to \(75\) (per meta-feature source) in
applications with \textless{} 10,000 tumors and \textless{} 20 cancer
sites. However, in applications with a larger number of tumors and/or
cancer sites, a larger \(S\) may be required for adequate model fit.

\hypertarget{sec:mcmc-details}{%
\section{Details for the MCMC sampler for posterior sampling from the
full model}\label{sec:mcmc-details}}

This section describes an MCMC sampling scheme for approximate sampling
from the model posterior distribution. At the outset, first note that
the normalized topic model parameters \(H\) and \(W\) are simple linear
transformations of the un-normalized topic model parameters
\(H = \tilde{H}D_{\tilde{W}}\) and \(W = D_{\tilde{W}}^{-1} \tilde{W}\)
where
\(D_{\tilde{W}} = \mathop{\mathrm{Diag}}(\sum_{p=1}^P \tilde{W}_{1p}, \dots, \sum_{p=1}^P \tilde{W}_{Sp})\).
This implies, \[
W^{-} = W^T(WW^T)^{-1} = \tilde{W}^T D_{\tilde{W}}^{-1} \left(D_{\tilde{W}}^{-1} \tilde{W}\tilde{W}^{T} D_{\tilde{W}}^{-1}\right)^{-1} = \tilde{W}(\tilde{W}\tilde{W}^T)^{-1} D_{\tilde{W}} = \tilde{W}^{-} D_{\tilde{W}}, 
\] and hence \(XUW^{-} = VW^{-} = V \tilde{W}^{-} D_{\tilde{W}}\)
(arises in the multinomial logistic likelihood). Therefore, for
posterior sampling we may focus on the unnormalized topic model
parameters \(\tilde{H}\) and \(\tilde{W}\) entirely along with the other
model parameters, and derive the the normalized topic model parameters
\(H\) and \(W\) posthoc via transformations. The joint posterior density
of all model parameters given the observed data is as follows: \[
\begin{aligned}
    & \pi(\alpha, \beta^0, \theta, \lambda^2, \{\tau_l^2\}, \{\tilde{H}_{ns}\},\{\tilde{W}_{sp}\}, \mid X, XU, c) \\ 
    &\propto  \prod_{n=1}^N \left( \frac{ \exp \left[ \alpha_{c_n} + \frac{e_{n;N}^T}{M_n}X\beta^0_{\bullet, c_n} + \frac{e_{n;N}^T}{M_n} V\tilde{W}^{-} D_{\tilde{W}} \theta_{\bullet,c_n} \right] }{\sum_{k'=1}^K \exp \left[\alpha_{k'} + \frac{e_{n;N}^T}{M_n}X\beta^0_{\bullet, k'} + \frac{e_{n;N}^T}{M_n} V \tilde{W}^{-} D_{\tilde{W}} \theta_{\bullet,k'}\right] }  \right) \\
    &\quad \times \prod_{s=1}^S \prod_{k=1}^K \left\{ \sigma_{\text{topic}, s} \ \exp\left(- \frac{\theta_{sk}^2 \sigma_{\text{topic}, s}^2}{2 \tau^2_{s;\theta}} \right) \right\} 
    \times \prod_{j=1}^J \prod_{k=1}^K \left\{ \sigma_{\text{obs}, j} \ \exp\left(- \frac{(\beta^0_{jk})^2 \sigma_{\text{obs}, j}^2}{2 \tau^2_{j;\beta^0}} \right) \right\} \\
    &\quad \times \prod_{j=1}^J \left(\lambda^2\right)^{\frac{K+1}{2}} \left(\tau^2_{j;\beta^0}\right)^{\frac{K+1}{2} - 1} \exp\left( - \frac{\lambda^2}{2} \tau^2_{j;\beta^0} \right)  
    \times \prod_{s=1}^S \left(\lambda^2\right)^{\frac{K+1}{2}} \left(\tau^2_{s;\theta}\right)^{\frac{K+1}{2} - 1} \exp\left( - \frac{\lambda^2}{2} \tau^2_{s;\theta} \right)  \\
     &\quad \times (\lambda^2)^{a_\lambda-1} \ e^{-b_\lambda \lambda^2} \\
    &\quad \times \prod_{n=1}^N \prod_{s=1}^S \prod_{p=1}^P \left( e^{-\tilde{H}_{ns}\tilde{W}_{sp}} \ \left(\tilde{H}_{ns} \tilde{W}_{sp}\right)^{Z_{nsp}} \right) \\
    &\quad \times \prod_{n=1}^N \prod_{s=1}^S \left(\tilde{H}_{ns}^{a_H - 1} \ e^{-b_H \tilde{H}_{ns}} \right) \times \prod_{s=1}^S \prod_{p=1}^P \left(\tilde{W}_{sp}^{a_W - 1} \ e^{-b_W \tilde{W}_{sp}} \right).
\end{aligned}
\]

Now, consider the transformation:
\({{\tilde{\theta}}}_{sk} = \sigma_{\text{topic}, s} \theta_{sk}\) and
\({{\tilde{\beta}}^0}_{jk} = \sigma_{\text{obs}, j} \beta^0_{jk}\). The
Jacobians of these transformations are \(1/\sigma_{\text{topic}, s}\)
and \(1/\sigma_{\text{obs}, j}\) respectively. In the matrix form, the
transformations can be written as
\({\tilde{\theta}}_{\bullet, k} = D_{\sigma_{\text{topic}}} \theta_{\bullet, k}\)
and
\({\tilde{\beta}}_{\bullet, k} = D_{\sigma_{\text{obs}}} \beta_{\bullet, k}\)
where
\(D_{\sigma_{\text{topic}}} = \mathop{\mathrm{Diag}}\left(\sigma_{\text{topic}, 1}, \dots, \sigma_{\text{topic}, S}\right)\)
and
\(D_{\sigma_{\text{obs}}} = \mathop{\mathrm{Diag}}\left(\sigma_{\text{obs}, 1}, \dots, \sigma_{\text{obs}, J}\right)\).
We shall again consider (MCMC) sampling for these transformed parameters
\({\tilde{\beta}}^0\) and \({\tilde{\theta}}\); the original parameters
\(\beta^0\) and \(\theta\) are computed posthoc via transformation. The
joint posterior density for \(\alpha\), \({\tilde{\beta}}^0\),
\({\tilde{\theta}}\), \(\lambda^2\), \(\{\tau_l^2\}\),
\(\{\tilde{H}_{ns}\}\), and \(\{\tilde{W}_{sp}\}\) is:

\begin{align*}
    & \pi(\alpha, \betat^0, \thetat, \lambda^2, \{\tau_l^2\}, \{\Ht_{ns}\},\{\Wt_{sp}\}, \mid X, XU, c) \\ 
    &\propto  \prod_{n=1}^N \left( \frac{ \exp \left[ \alpha_{c_n} + \frac{e_{n;N}^T}{M_n} X D_{\sigma_{\text{obs}}}^{-1} \betat^0_{\bullet, c_n} + \frac{e_{n;N}^T}{M_n} V\Wt^{-} D_{\Wt} D_{\sigma_{\text{topic}}}^{-1} \thetat_{\bullet,c_n} \right] }{\sum_{k'=1}^K \exp \left[\alpha_{k'} + \frac{e_{n;N}^T}{M_n} X D_{\sigma_{\text{obs}}}^{-1} \betat^0_{\bullet, k'} + \frac{e_{n;N}^T}{M_n} V \Wt^{-} D_{\Wt} D_{\sigma_{\text{topic}}}^{-1} \thetat_{\bullet,k'}\right] }  \right) \\
    &\quad \times \prod_{s=1}^S \prod_{k=1}^K \left\{ \ \exp\left(- \frac{\thetat_{sk}^2 }{2 \tau^2_{s;\theta}} \right) \right\} \times \prod_{j=1}^J \prod_{k=1}^K \left\{ \ \exp\left(- \frac{(\betat^0_{jk})^2 }{2 \tau^2_{j;\beta^0}} \right) \right\} \\
    &\quad \times \prod_{j=1}^J \left(\lambda^2\right)^{\frac{K+1}{2}} \left(\tau^2_{j;\beta^0}\right)^{\frac{K+1}{2} - 1} \exp\left( - \frac{\lambda^2}{2} \tau^2_{j;\beta^0} \right) \times \prod_{s=1}^S \left(\lambda^2\right)^{\frac{K+1}{2}} \left(\tau^2_{s;\theta}\right)^{\frac{K+1}{2} - 1} \exp\left( - \frac{\lambda^2}{2} \tau^2_{s;\theta} \right)  \\
     &\quad \times (\lambda^2)^{a_\lambda-1} \ e^{-b_\lambda \lambda^2} \\
    &\quad \times \prod_{n=1}^N \prod_{s=1}^S \prod_{p=1}^P \left( e^{-\Ht_{ns}\Wt_{sp}} \ \left(\Ht_{ns} \Wt_{sp}\right)^{Z_{nsp}} \right) \\
    &\quad \times \prod_{n=1}^N \prod_{s=1}^S \left(\Ht_{ns}^{a_H - 1} \ e^{-b_H \Ht_{ns}} \right) \times \prod_{s=1}^S \prod_{p=1}^P \left(\Wt_{sp}^{a_W - 1} \ e^{-b_W \Wt_{sp}} \right).
\end{align*}

Below we develop an approximate collapsed Metropolis-within-Gibbs
sampler for MCMC sampling from the above posterior. The sampler
iteratively draws random elements (MCMC samples) from (a) the
conditional posterior
\(\pi(\tilde{H}, \tilde{W}\mid \alpha, {\tilde{\beta}}^0, {\tilde{\theta}}, X, V, c)\)
for the \emph{non-normalized} topic model parameters \(\tilde{H}\),
\(\tilde{W}\) given the multinomial logistic regression (supervised)
parameters, and (b) from the conditional posterior
\(\pi(\alpha, {\tilde{\beta}}^0, {\tilde{\theta}}, \lambda^2, \{\tau_l^2\} \mid \tilde{H}, \tilde{W}, X, V, c)\)
of the multinomial logistic regression given the topic model parameter
induced predictor matrices.

\hypertarget{approximate-sampling-from-the-supervised-conditional-topic-model-posterior-pitildeh-tildewmid-alpha-tildebeta0-tildetheta-lambda2-tau_l2-x-v-c}{%
\subsection{\texorpdfstring{Approximate sampling from the supervised
conditional topic model posterior
\(\pi(\tilde{H}, \tilde{W}\mid \alpha, {\tilde{\beta}}^0, {\tilde{\theta}}, \lambda^2, \{\tau_l^2\}, X, V, c)\)}{Approximate sampling from the supervised conditional topic model posterior \textbackslash pi(\textbackslash tilde\{H\}, \textbackslash tilde\{W\}\textbackslash mid \textbackslash alpha, \{\textbackslash tilde\{\textbackslash beta\}\}\^{}0, \{\textbackslash tilde\{\textbackslash theta\}\}, \textbackslash lambda\^{}2, \textbackslash\{\textbackslash tau\_l\^{}2\textbackslash\}, X, V, c)}}\label{approximate-sampling-from-the-supervised-conditional-topic-model-posterior-pitildeh-tildewmid-alpha-tildebeta0-tildetheta-lambda2-tau_l2-x-v-c}}

Exploiting the additive property of the Poisson distribution and the
multinomial-Poisson connection, we device a MCMC sampling strategy from
the conditional posterior of the supervised conditional topic model.
Subsequently, approximate samples from the posterior
\(\pi(\tilde{H}, \tilde{W}, Z \mid \alpha, {\tilde{\beta}}^0, {\tilde{\theta}}, \lambda^2, \{\tau_l^2\}, X, V, c)\)
is generated iteratively by first drawing random latent data \(Z\) from
the conditional density
\(\pi(Z \mid \tilde{H}, \tilde{W}, \alpha, {\tilde{\beta}}^0, {\tilde{\theta}}, X, V, c) = \pi(Z \mid \tilde{H}, \tilde{W}, V)\),
and then drawing Markov chain draws for \((\tilde{H}, \tilde{W})\) from
the conditional density
\(\pi(\tilde{H}, \tilde{W}\mid Z, \alpha, {\tilde{\beta}}^0, {\tilde{\theta}}, X, V, c)\).

\hypertarget{drawing-z_nsp}{%
\subsubsection{\texorpdfstring{Drawing
\(\{Z_{nsp}\}\)}{Drawing \textbackslash\{Z\_\{nsp\}\textbackslash\}}}\label{drawing-z_nsp}}

Conditional on \(V_{np} = \sum_{s = 1}^S Z_{nsp}\), the vector
\((Z_{n1p}, \dots, Z_{nSp})\) has a multinomial distribution: \[
  (Z_{n1p}, \dots, Z_{nSp}) \mid \tilde{H}, \tilde{W}, V \sim \text{Multinomial}(V_{np}, (\phi_{n1p}, \dots, \phi_{nSp})) 
\] where \(\phi_{nsp} \propto \tilde{H}_{ns} \tilde{W}_{sp}\) and
\(\sum_{s=1}^S \phi_{nsp} = 1\), and the collection \(\{Z_{nsp}\}\) is
independent of \(\{Z_{n's'p'}\}\) whenever \(n \neq n'\) or
\(p \neq p'\).

\hypertarget{drawing-tildeh-tildew-approximately}{%
\subsubsection{\texorpdfstring{Drawing \(\tilde{H}, \tilde{W}\)
(approximately)}{Drawing \textbackslash tilde\{H\}, \textbackslash tilde\{W\} (approximately)}}\label{drawing-tildeh-tildew-approximately}}

The full conditional density of \(\{\tilde{H}_{ns}\}\),
\(\{\tilde{W}_{sp}\}\) given the latent data \(\{Z_{nsp}\}\), all other
parameters and the observed data is given by:

\[
    \begin{aligned}
    & \pi(\{\tilde{H}_{ns}\},\{\tilde{W}_{sp}\},  \mid \{Z_{nsp}\}, \alpha, {\tilde{\beta}}^0,  {\tilde{\theta}}, \lambda^2, \{\tau_l^2\}, X, V, c) \\ 
    & \propto \prod_{n=1}^N \prod_{s=1}^S \left(\tilde{H}_{ns}^{a_H - 1} \ e^{-b_H \tilde{H}_{ns}} \right) \times \prod_{s=1}^S \prod_{p=1}^P \left(\tilde{W}_{sp}^{a_W - 1} \ e^{-b_W \tilde{W}_{sp}} \right) \\
    &\quad \times \prod_{n=1}^N \prod_{s=1}^S \prod_{p=1}^P \left( e^{-\tilde{H}_{ns}\tilde{W}_{sp}} \ \left(\tilde{H}_{ns} \tilde{W}_{sp}\right)^{Z_{nsp}} \right) \\
    &\quad \times \prod_{n=1}^N \left( \frac{ \exp \left[ \alpha_{c_n} + \frac{e_{n;N}^T}{M_n} X D_{\sigma_{\text{obs}}}^{-1} \beta^0_{\bullet, c_n} + \frac{e_{n;N}^T}{M_n} \sum_{s=1}^S \left(V \tilde{W}^{-} D_{\tilde{W}}\right)_{\bullet, s} \frac{{\tilde{\theta}}_{s,c_n}}{\sigma_{\text{topic}, s}} \right] }{\sum_{k'=1}^K \exp \left[\alpha_{k'} + \frac{e_{n;N}^T}{M_n} X D_{\sigma_{\text{obs}}}^{-1} \beta^0_{\bullet, k'} + \frac{e_{n;N}^T}{M_n} \sum_{s=1}^S \left(V \tilde{W}^{-} D_{\tilde{W}}\right)_{\bullet, s} \frac{{\tilde{\theta}}_{s,k'}}{\sigma_{\text{topic}, s}}\right] }  \right).
    \end{aligned}
\]

To generate approximate samples from the above density a further Gibbs
type algorithm will be devised to draw \(\tilde{H}\) and \(\tilde{W}\)
from their corresponding conditional densities.

\quad We first discuss an \emph{exact} Metropolis-within-Gibbs sampler
to draw from the above conditional density, and note the computational
challenges associated with this sampler. In this sampler, one
iteratively draws \(\{\tilde{H}_{ns}\}\) from the conditional posterior
density given \(\{Z_{nsp}\}\), \(\tilde{W}\) and the multinomial
logistic parameters, and then draws \(\{\tilde{W}_{sp}\}\) from its
conditional density given \(\tilde{H}\), \(\{Z_{nsp}\}\) and the
multinomial logistic parameters. The conditional density of
\(\{\tilde{H}_{ns}\}\) is product gamma which can be effortlessly
simulated. The (joint) conditional density of \(\{\tilde{W}_{sp}\}\)
however is not in a standard form, and has the shape of a product gamma
density (contribution from the unsupervised part) multiplied with the
product of multinomial logistic likelihood terms (supervised part). In
theory one may consider a Metropolis-Hastings MCMC sampling scheme here,
by first drawing \(\{\tilde{W}_{sp}\}\) proposals from the product Gamma
density arising from the unsupervised part, and then performing a
Metropolis accept-reject step to account for the contribution of
supervised part. A key computational challenge with this approach lies
in the evaluation of the generalized inverse matrix \(\tilde{W}^{-}\)
which must be evaluated for each proposal of the \(\tilde{W}\) matrix
which adds substantial computational burden to the entire process.
Furthermore, the contribution to the supervised multinomial logistic
likelihood of individual data points \(\{n\}\) cannot be separated as
they all include contributions of the entire \(\tilde{W}\) matrix. The
collective (product) logistic contributions of all data points however
may substantially lower the Metropolis acceptance rate.

Instead, we develop and use the following following \emph{approximate}
Metropolis-within-Gibbs sampler for sampling from
\(\tilde{H}, \tilde{W}\), by invoking the following approximations.

\textbf{Approximation 1:}

First, we approximate \(Z_{nsp}\) in the multinomial logistic term in
the above joint density by its conditional expectation
\(Z_{nsp} \approx E(Z_{nsp} \mid \tilde{H}_{ns}, \tilde{W}_{nsp}) = \tilde{H}_{nsp} \tilde{W}_{sp}\).
This is the usual NMF approximation, and can be viewed as a first order
Taylor approximation. Note that this implies
\(V_{np} = \sum_{s=1}^S Z_{nsp} \approx E(\sum_{s=1}^S Z_{nsp} \mid \tilde{H}_{ns}, \tilde{W}_{nsp}) = \sum_{s=1}^S \tilde{H}_{nsp} \tilde{W}_{sp} = (\tilde{H}\tilde{W})_{sp}\),
i.e., \(V = XU \approx \tilde{H}\tilde{W}\). Then for each
\(k = 1, \dots, K\),

\[
\begin{aligned}
\frac{e_{n;N}^T}{M_n} \sum_{s=1}^S \left(V \tilde{W}^{-} D_{\tilde{W}}\right)_{\bullet, s} \frac{{\tilde{\theta}}_{s,k}}{\sigma_{\text{topic}, s}} 
&\approx \frac{e_{n;N}^T}{M_n} \sum_{s=1}^S \left(\tilde{H}\tilde{W}\tilde{W}^{-} D_{\tilde{W}}\right)_{\bullet, s} \frac{{\tilde{\theta}}_{s,k}}{\sigma_{\text{topic}, s}} \\
&= \frac{e_{n;N}^T}{M_n} \sum_{s=1}^S \left(\tilde{H}D_{\tilde{W}}\right)_{\bullet, s} \frac{{\tilde{\theta}}_{s,k}}{\sigma_{\text{topic}, s}} \\
&= \frac{e_{n;N}^T}{M_n} \sum_{s=1}^S \left[\tilde{H}\right]_{\bullet, s} \left[D_{\tilde{W}}\right]_{s,s} \frac{{\tilde{\theta}}_{s,k}}{\sigma_{\text{topic}, s}} \\
&= \frac{1}{M_n} \tilde{H}_{ns} \left[D_{\tilde{W}}\right]_{s,s} \frac{{\tilde{\theta}}_{s,k}}{\sigma_{\text{topic}, s}} 
\end{aligned}
\] and \[
\begin{aligned}
\sigma_{\text{topic}, s} 
&= g\left(\left[\tilde X U \tilde{W}^{-} D_{\tilde{W}}\right]_{\bullet, s}\right)  \\
&= g\left(\left[D_{M}^{-1} X U  \tilde{W}^{-} D_{\tilde{W}}\right]_{\bullet, s}\right) \\
&\approx  g\left(\left[D_{M}^{-1} \tilde{H}D_{\tilde{W}} \right]_{\bullet, s} \right) \\
&= \left[D_{\tilde{W}}\right]_{s,s} \  g\left(D_{M}^{-1} \left[\tilde{H}\right]_{\bullet, s} \right)
\end{aligned}
\] where \(\left[D_{\tilde{W}}\right]_{s,s} = \sum_{p=1}^P W_{sp}\) is
the \(s\)-th diagonal entry of \(D_{\tilde{W}}\),
\(\left[\tilde{H}\right]_{\bullet, s}\) is the \(s\)-th column of
\(\tilde{H}\), \(D_{M} = \text{Diag}(M_1, \dots, M_n)\), and \(g\) is
the sample standard deviation function defined for a vector
\(t = (t_1, \dots, t_N)\) as
\(g(t) = \sqrt{\frac{1}{N-1} \sum_{n=1}^N \left(t_n - \bar t \right)^2}\)
with \(\bar t = \frac1N \sum_{n=1}^N t_n\). Combining, we get \[
\frac{e_{n;N}^T}{M_n} \sum_{s=1}^S \left(V \tilde{W}^{-} D_{\tilde{W}}\right)_{\bullet, s} \frac{{\tilde{\theta}}_{s,k}}{\sigma_{\text{topic}, s}}  \approx \sum_{s=1}^S \tilde{H}_{ns} \frac{{\tilde{\theta}}_{s,k}}{M_n \ g\left(D_{M}^{-1} \left[\tilde{H}\right]_{\bullet, s} \right)}
\] which implies

\begin{align*}
&\pi(\{\Ht_{ns}\},\{\Wt_{sp}\} \mid \alpha, \betat^0, \thetat, \lambda^2, \{\tau_l^2\}, X, V, c)  \\ 
& \underset{\sim}{\propto} \prod_{n=1}^N \prod_{s=1}^S \left(\Ht_{ns}^{a_H - 1} \ e^{-b_H \Ht_{ns}} \right) \times \prod_{s=1}^S \prod_{p=1}^P \left(\Wt_{sp}^{a_W - 1} \ e^{-b_W \Wt_{sp}} \right) \\
&\quad \times \prod_{n=1}^N \prod_{s=1}^S \prod_{p=1}^P \left( e^{-\Ht_{ns}\Wt_{sp}} \ \left(\Ht_{ns} \Wt_{sp}\right)^{Z_{nsp}} \right) \\
&\quad \times \prod_{n=1}^N \left( \frac{ \exp \left[ \alpha_{c_n} + \frac{e_{n;N}^T}{M_n} X D_{\sigma_{\text{obs}}}^{-1} \betat^0_{\bullet, c_n} + \sum_{s=1}^S \Ht_{ns} \frac{\thetat_{s, c_n}}{M_n \ g\left(D_{M}^{-1} \left[\Ht \right]_{\bullet, s} \right)} \right] }{\sum_{k'=1}^K \exp \left[\alpha_{k'} +  \frac{e_{n;N}^T}{M_n} X D_{\sigma_{\text{obs}}}^{-1} \betat^0_{\bullet, k'} + \sum_{s=1}^S \Ht_{ns} \frac{\thetat_{s, k'}}{M_n \ g\left(D_{M}^{-1} \left[\Ht \right]_{\bullet, s} \right)} \right] }  \right) .
\end{align*}

There are two major consequences of this approximation. First, the
approximate joint conditional posterior density does not involve the
generalized inverse matrix \(\tilde{W}^{-}\) any more, which effectuates
significant reduction in the overall computation burden. Second, the
full conditional (given \(\{\tilde{H}_{ns}\}\) and other parameters)
joint posterior density of \(\{\tilde{W}_{sp}\}\) reduces to a product
univariate gamma form: \[
\begin{aligned}
\pi(\{\tilde{W}_{sp}\} \mid \{\tilde{H}_{ns}\}, \alpha, \beta^0, \theta, \lambda^2, \{\tau_l^2\}, X, V, c)  \propto  \prod_{s=1}^S \prod_{p=1}^P \left(\tilde{W}_{sp}^{a_W + \sum_{n=1}^N Z_{nsp} - 1} \ e^{-(b_W + \sum_{n=1}^N \tilde{H}_{ns}) \tilde{W}_{sp}} \right).
\end{aligned}
\] This implies that conditional on the other parameters, the entries of
the \(\tilde{W}\) matrix can be effortlessly and independently sampled
from univariate gamma distributions, which leads to the following step
for the MCMC sampling.

\begin{enumerate}
\def\labelenumi{\arabic{enumi}.}
\tightlist
\item
  \textbf{Drawing} \(\tilde{W}\)\textbf{.} Generate
  \(\{\tilde{W}_{sp}\}\) independently from the following full
  conditional Gamma distribution: \[
  \tilde{W}_{sp} \mid Z, \tilde{H}, \alpha, \beta^0, \theta, \lambda^2, \{\tau_l^2\}, X, V, c \overset{a}{\sim} \text{Gamma}\left(a_W + \sum_{n=1}^N Z_{nsp}, b_W + \sum_{n=1}^N \tilde{H}_{ns}\right).
  \]
\end{enumerate}

However, the approximate full conditional joint density of
\(\{\tilde{H}_{ns}\}\) given \(\{\tilde{W}_{sp}\}\) and other parameters
still has an intractable form due to the softmax (multinomial)
probabilities which also include the computed sample standard deviation
terms. Note that, for each topic \(s\), the approximate sample standard
deviation term
\(g\left(D_{M}^{-1} \left[\tilde{H}\right]_{\bullet, s} \right)\)
involves all rows of the matrix column \(\tilde{H}_{\bullet, s}\)
arising out of different \(\{n\}\), and the multinomial logistic
contribution of each observation \(n\) involves all topics \(s\). In
other words, the elements of \(\tilde{H}\) cannot be independently
sampled, even when conditioned on \(\tilde{W}\) and \(Z\). We propose
the following \emph{independent} Metropolis-Hastings sampler that
generates each row of \(\tilde{H}\) sequentially conditional on all
other rows of \(\tilde{H}\). An approximation that reduces the
computational load and encourages better mixing by removing the
dependence across rows is discussed afterward.

\begin{enumerate}
\def\labelenumi{\arabic{enumi}.}
\setcounter{enumi}{1}
\tightlist
\item
  \textbf{Drawing the rows of} \(\tilde{H}\). Let
  \(\tilde{H}^{\text{old}}\) denote the current state of \(\tilde{H}\).
  Each row of \(\tilde{H}\) is drawn sequentially conditional on the
  other rows. To update the \(n\)-th row first draw a proposal vector
  \(\tilde{H}_{n, \bullet}^{\text{prop}} = (\tilde{H}_{n1}^{\text{prop}}, \dots, \tilde{H}_{nS}^{\text{prop}})\)
  utilizing the independent unsupervised Poisson contribution for the
  each augmented count \(\{Z_{nsp}\}\): \[
  \tilde{H}_{ns}^{\text{prop}} \mid Z, \tilde{W}, X, V \overset{a}{\sim} \text{Gamma}\left( a_H + \sum_{p=1}^P Z_{nsp}, b_H + \sum_{p=1}^P \tilde{W}_{sp} \right).
  \] With these proposal values \(\{\tilde{H}_{ns}^{\text{prop}}\}\) for
  the \(n\)-th row and keeping the other rows fixed at their current
  states, compute (update) the sample standard deviation terms
  \(\tilde \sigma_{s} = \tilde \sigma_{s}([\tilde{H}]_{\bullet, s}) = g\left(D_{M}^{-1} \left[\tilde{H}\right]_{\bullet, s} \right)\).
  Call the corresponding (proposal) value of \(\tilde \sigma_s\) as
  \(\tilde \sigma^{\text{prop}}_s\) and the current value of
  \(\tilde \sigma_s\) as \(\sigma^{\text{prop}}_s\). Next compute the
  acceptance ratio \[
  \begin{aligned}
  r\left(\tilde{H}_{n, \bullet}^{\text{prop}} \mid \tilde{H}_{n, \bullet}^{\text{old}}\right) 
  &= \left( \frac{ \exp \left[ \alpha_{c_n} + \frac{e_{n;N}^T}{M_n} X D_{\sigma_{\text{obs}}}^{-1} {\tilde{\beta}}^0_{\bullet, c_n} + \sum_{s=1}^S {\tilde{H}}^{\text{prop}}_{ns} \frac{{\tilde{\theta}}_{s, c_n}}{M_n \ \tilde \sigma^{\text{prop}}_s } \right] }{\sum_{k'=1}^K \exp \left[\alpha_{k'} +  \frac{e_{n;N}^T}{M_n} X D_{\sigma_{\text{obs}}}^{-1} {\tilde{\beta}}^0_{\bullet, k'} + \sum_{s=1}^S {\tilde{H}}^{\text{prop}}_{ns} \frac{{\tilde{\theta}}_{s, k'}}{M_n \ \sigma^{\text{prop}}_s} \right] }  \right) \\
  &\quad \times \left( \frac{ \exp \left[ \alpha_{c_n} + \frac{e_{n;N}^T}{M_n} X D_{\sigma_{\text{obs}}}^{-1} {\tilde{\beta}}^0_{\bullet, c_n} + \sum_{s=1}^S {\tilde{H}}^{\text{old}}_{ns} \frac{{\tilde{\theta}}_{s, c_n}}{M_n \ \tilde \sigma^{\text{old}}_s } \right] }{\sum_{k'=1}^K \exp \left[\alpha_{k'} +  \frac{e_{n;N}^T}{M_n} X D_{\sigma_{\text{obs}}}^{-1} {\tilde{\beta}}^0_{\bullet, k'} + \sum_{s=1}^S {\tilde{H}}^{\text{old}}_{ns} \frac{{\tilde{\theta}}_{s, k'}}{M_n \ \sigma^{\text{old}}_s} \right] }  \right)^{-1}
  \end{aligned}    
  \]
\end{enumerate}

and draw \(U \sim \text{Uniform}(0, 1)\). Then update the \(n\)-th row
of \(\tilde{H}\) by \(\tilde{H}_{n, \bullet}^{\text{prop}}\) if
\(U < r(\tilde{H}_{n, \bullet}^{\text{prop}} \mid \tilde{H}_{n, \bullet}^{\text{old}})\)
or leave the row unchanged otherwise. We note that this one-step
Metropolis adjustment for the entire \(n\)-th row for \(\tilde{H}\) may
have low acceptance rate if the choices of the regression coefficient
\(\{\theta_{sp}\}\) are poor, and/or the total number of topics \(S\) is
large. For large \(S\) (say \(S \geq 50\)) we recommend partitioning
each row into multiple smaller dimensional random blocks and then
performing the above independent Metropolis-Hastings sampling separately
on each block conditioning the remaining blocks. Simultaneously, the
Metropolis accept-reject step for each row proposal may be performed
multiple times with independent, new proposals, until a reasonable (say
90\%\footnote{Note that the proposed Metropolis sampler is an independence sampler, and not a random-walk Metropolis algorithm. Unlike random-walk Metropolis algorithms, independence Metropolis samplers do not require manual tuning, and can achieve better convergence than random walk samplers depending on the problem.})
overall acceptance rate is achieved.

\textbf{Approximation 2:}

The serial dependence among the rows of \(\tilde{H}\) and the overall
computational burden can be reduced by assuming
\(\tilde \sigma_{s}([\tilde{H}]_{\bullet, s}) = g\left(D_{M}^{-1} \left[\tilde{H}\right]_{\bullet, s} \right)\)
to be approximately fixed across iterations, say at
\(\tilde \sigma_{s}^{\text{fixed}}\). This can be a reasonable
approximation when the sample size \(N\) is large and the Markov chain
is stationary; in which case the
\(\tilde \sigma_{s}([\tilde{H}]_{\bullet, s}) \approx \sigma_{s}^{\text{popn}}\).
Here \(\sigma_{s}^{\text{popn}}\) denotes the population (associated
with the tumor samples \(\{n\}\)) level standard deviation of the
exposure from the \(s\)-th topic (cataloged along the rows of
\(D_{M}^{-1} \left[\tilde{H}\right]_{\bullet, s}\)). Then, the
Metropolis acceptance ratio becomes \[
\begin{aligned}
\tilde r\left(\tilde{H}_{n, \bullet}^{\text{prop}} \mid \tilde{H}_{n, \bullet}^{\text{old}}\right) 
&= \left( \frac{ \exp \left[ \alpha_{c_n} + \frac{e_{n;N}^T}{M_n} X D_{\sigma_{\text{obs}}}^{-1} {\tilde{\beta}}^0_{\bullet, c_n} + \sum_{s=1}^S {\tilde{H}}^{\text{prop}}_{ns} \frac{{\tilde{\theta}}_{s, c_n}}{M_n \ \tilde \sigma^{\text{fixed}}_s } \right] }{\sum_{k'=1}^K \exp \left[\alpha_{k'} +  \frac{e_{n;N}^T}{M_n} X D_{\sigma_{\text{obs}}}^{-1} {\tilde{\beta}}^0_{\bullet, k'} + \sum_{s=1}^S {\tilde{H}}^{\text{prop}}_{ns} \frac{{\tilde{\theta}}_{s, k'}}{M_n \ \sigma^{\text{fixed}}_s} \right] }  \right) \\
&\quad \times \left( \frac{ \exp \left[ \alpha_{c_n} + \frac{e_{n;N}^T}{M_n} X D_{\sigma_{\text{obs}}}^{-1} {\tilde{\beta}}^0_{\bullet, c_n} + \sum_{s=1}^S {\tilde{H}}^{\text{old}}_{ns} \frac{{\tilde{\theta}}_{s, c_n}}{M_n \ \tilde \sigma^{\text{fixed}}_s } \right] }{\sum_{k'=1}^K \exp \left[\alpha_{k'} +  \frac{e_{n;N}^T}{M_n} X D_{\sigma_{\text{obs}}}^{-1} {\tilde{\beta}}^0_{\bullet, k'} + \sum_{s=1}^S {\tilde{H}}^{\text{old}}_{ns} \frac{{\tilde{\theta}}_{s, k'}}{M_n \ \sigma^{\text{fixed}}_s} \right] }  \right)^{-1}
\end{aligned}    
\] Consequently, under this approximation, the Metropolis steps for each
row of \(\tilde{H}\) can be performed independently, permitting better
mixing, and reduction in the computation load (due to iterative updating
of the standard deviation terms), at the cost of some (possibly small)
approximation error.

\hypertarget{sampling-from-the-conditional-multinomial-logistic-model-posterior-given-topic-model-parameters}{%
\subsection{Sampling from the conditional multinomial logistic model
posterior given topic model
parameters}\label{sampling-from-the-conditional-multinomial-logistic-model-posterior-given-topic-model-parameters}}

We develop the following data augmentation Gibbs sampler for MCMC
sampling from the conditional, group-lasso regularized, multinomial
logistic posterior given the topic model parameters. Conditional on the
topic model, the multinomial logistic regression entails predicting the
cancer site probabilities \(\{\Pr(c = k)\}\) given the topic model
induced \emph{row-normalized} and \emph{column standard-deviation
scaled} stacked design matrix
\((1, D_M^{-1} X D_{\sigma_{\text{obs}}}^{-1}, D_M^{-1} XUW^{-} D_{\sigma_{\text{topic}}}^{-1})\).
Let \(L = (S + J)\), and construct the collected regression coefficient
matrix \(\delta\) by stacking the rows of \(\beta^0\) and \(\theta\)
matrices; the \(k\)-th column of \(\delta\) is thus given by \[
\delta_{\bullet, k} = 
\begin{pmatrix}
{\tilde{\beta}}^0_{1, k} \\
\vdots \\
{\tilde{\beta}}^0_{J, k} \\
{\tilde{\theta}}_{1, k} \\
\vdots \\
{\tilde{\theta}}_{S, k} \\
\end{pmatrix}.
\] We index the rows of \(\delta\) by \(l; l = 1, \dots, L\). The
proposed Gibbs sampler utilizes the Pólya-Gamma data augmentation scheme
for (multinomial) logistic regression model (Polson, Scott, and Windle
2013) and the Gibbs sampler for hierarchical group-lasso regularized
linear regression (Casella et al. 2010). At each iteration, the sampler
iterates between drawing the sparsity handling variance parameters
\(\{\tau_l^2\}\) and \(\lambda^2\), and the multinomial logistic
parameters \(\alpha\), and \(\delta\).

\hypertarget{drawing-the-sparsity-handling-parameters}{%
\subsubsection{Drawing the sparsity handling
parameters}\label{drawing-the-sparsity-handling-parameters}}

\begin{enumerate}
\def\labelenumi{\arabic{enumi}.}
\item
  \textbf{Drawing} \(\{\tau_l^2\}\). Draw the reciprocals
  \(\{1/\tau_l^2: l = 1, \dots, L\}\) independently from the following
  conditional posterior distributions: \[ 
  1/\tau_l^2 \mid \{\delta\}, \lambda \sim 
  \begin{cases}
  \text{inverse-Gamma}\left(\frac{1}{2}, \frac{\lambda^2}{2}  \right) & \|\delta_{l, \bullet}\| = 0 \\
  \text{inverse-Gaussian}\left( \sqrt{\frac{\lambda^2}{\|\delta_{l, \bullet}\|^2}}, \lambda^2 \right)  & \|\delta_{l, \bullet}\| > 0 
  \end{cases}
  \] and subsequently compute \(\{\tau_l^2: l = 1, \dots, D\}\).
\item
  \textbf{Drawing} \(\lambda^2\): Generate \(\lambda^2\) from its
  conditional posterior density given \({\tau_l^2}\): \[
  \lambda^2 \mid \{\tau_l^2\} \sim \text{Gamma}\left( a_\lambda + \frac{K(L + 1)}{2}, b_\lambda + \frac{\sum_{l=1}^{L} \tau_l^2}{2} \right).
  \]
\end{enumerate}

\hypertarget{drawing-augmented-data-and-multinomial-logistic-regression-parameters}{%
\subsubsection{Drawing augmented data and multinomial logistic
regression
parameters}\label{drawing-augmented-data-and-multinomial-logistic-regression-parameters}}

First define the derived conditional design matrix
\(\accentset{\approx}{X}\) by column stacking: \[
\accentset{\approx}{X} = \left( 1 \ \vdots \ D_M^{-1} X D_{\sigma_{\text{obs}}}^{-1} \ \vdots \ D_M^{-1} XUW^{-} D_{\sigma_{\text{topic}}}^{-1} \right)
\] and denote the rows of \(\accentset{\approx}{X}\) as
\(\accentset{\approx}{x}_n\); \(n = 1, \dots, N\). Next separately for
each cancer site \(k = 1, \dots, K\) iteratively update the associated
intercept and regression coefficients
\(\tilde{\delta}_k = (\alpha_k, \delta_{\bullet, k}^T)^T\) conditional
on those for the other cancer sites \(k' \neq k\) utilizing a
Pólya-Gamma data augmentation scheme (Polson, Scott, and Windle 2013) as
follows.

\begin{enumerate}
\def\labelenumi{\arabic{enumi}.}
\item
  \textbf{Generate Pólya-Gamma augmented data} \(\{\gamma_{nk}\}\).
  Calculate \[
  g_{nk} = \log \left( \sum_{k'\neq k} \accentset{\approx}{x}_n^T \tilde \delta_k \right); \ G_{nk} = \accentset{\approx}{x}_n^T \tilde \delta_k - g_{nk}  
  \] and draw the augmented data \(\{\gamma_{nk}: n = 1, \dots, k\}\)
  from the Pólya-Gamma distribution (Polson, Scott, and Windle 2013)\[
  \gamma_{nk} \mid \text{rest} \sim \text{PG}(1, g_{nk}).
  \] and define
  \(\Gamma_k = \text{Diag}(\gamma_{1k}, \dots, \gamma_{nk})\).
\item
  \textbf{Generate multinomial logistic intercept and regression
  parameters.} Calculate \(B_k\) and \(\mu_k\) as follows \[
  B_k^{-1} = \left({\accentset{\approx}{X}}^T \Gamma_k \ \accentset{\approx}{X} \right) + \text{Diag}(1/\sigma^2_{\alpha, 0k}, 1/\tau_1^2, \dots, 1/\tau_L^2)
  \] and \[
  \mu_k = B_k {\accentset{\approx}{X}}^T \left( \left[y_{\bullet, k} - \frac12 \right] + \Gamma_{k} \ C_{\bullet, k} \right)
  \] where \(y_{\bullet, k}\) is the \(k\)-th column of the binary
  cancer site indicator matrix \(((y_{nk}))\) with
  \(y_{nk} = I(c_n = k)\) and \(C_{\bullet, k}\) is the \(k\)-th column
  of the matrix \(((C_{nk}))\). The intercept \(\alpha_k\) and the
  regression coefficients \(\delta_{\bullet, k}\) are then jointly
  generated from the conditional multivariate normal distribution: \[
  \tilde \delta_k = (\alpha_k, \delta_{\bullet, k}^T)^T \mid \text{rest} \sim \text{N}_{1+L}(\mu_k, B_k)
  \]
\end{enumerate}

\hypertarget{initialization-of-the-mcmc-algorithm}{%
\section{Initialization of the MCMC
algorithm}\label{initialization-of-the-mcmc-algorithm}}

The MCMC algorithm is initialized at the approximate marginal posterior
mode of the model. To this end, we first obtain the marginal posterior
mode of the multi-logistic intercept \(\alpha\), regression coefficients
for the residual variant effects \(\beta^0\) and the \emph{observed}
meta-feature effects \(\omega\) using the group-lasso based point
estimation method of (Chakraborty, Begg, and Shen 2020) on the training
data using the row normalized versions of the mutation counts \(X\) and
\(XU\). The associated cross-validation based optimal value of the group
lasso penalty parameter \(\lambda\) is also noted. Next, we perform
frequentist unsupervised NMF estimation (DeBruine 2021) on the columns
of \(XU\) to obtain initial estimates of \(H\) and \(W\). The initial
values of the regression coefficients \(\theta\) for the \emph{latent
topics} are then obtained as \(\theta = W \omega\). These initial values
of the parameter \(\alpha\), \(\beta^0\), \(\theta\), \(H\) and \(W\)
are then used as then used to generate initial values of the remaining
parameters \(\{\tau_l^2\}\) based on the MCMC sampling scheme.

\pagebreak

\hypertarget{association-of-the-liver-window-topic-specific-composition-probabilities-figure-2a-in-the-main-text-with-h3k4me1-histone-marks}{%
\section{Association of the ``Liver'' window topic-specific composition
probabilities (Figure 2A in the main text) with H3K4me1 Histone
Marks}\label{association-of-the-liver-window-topic-specific-composition-probabilities-figure-2a-in-the-main-text-with-h3k4me1-histone-marks}}

\begin{figure}[!htb]
    \centering
    \includegraphics[width=\textwidth]{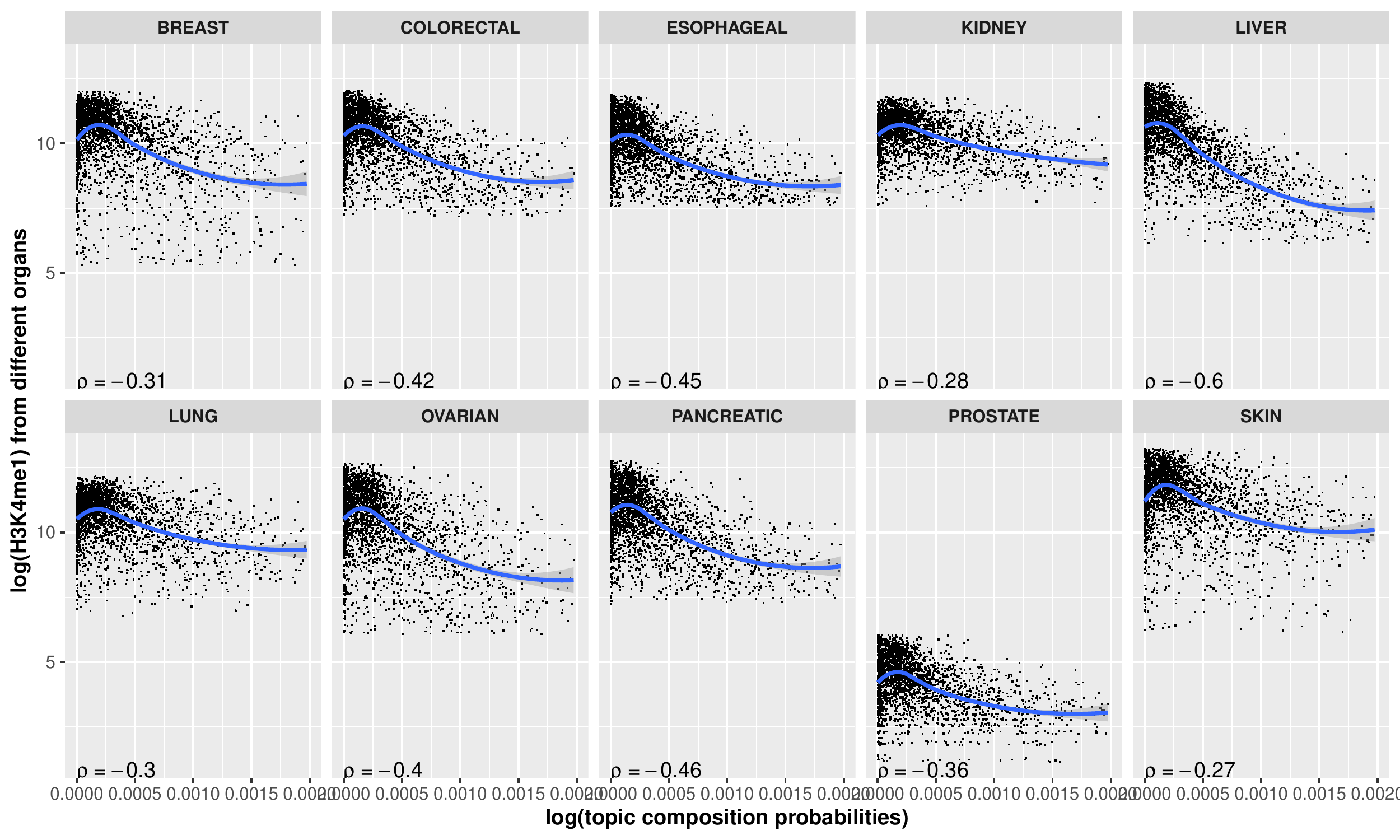}
    \caption{Points in the scatters display the log  composition probabilities for the "liver" topic (displayed on Figure 2A in the main text) along the x-axes and log epignomic features (histone marks H3k4me1) along the y-axes for all $\sim 3000$ chromosome window meta-feature categories. Different site-specific epigenome scores are plotted along different panels. Inscribed at the bottom-left corner of each panel is the Spearman correlation coefficient $\rho$ between the corresponding topic composition probabilities and H3k4me1 scores.}
    \label{fig:window-topic-epigenome}
\end{figure}

\pagebreak

\hypertarget{references}{%
\section*{References}\label{references}}
\addcontentsline{toc}{section}{References}

\hypertarget{refs}{}
\begin{CSLReferences}{1}{0}
\leavevmode\vadjust pre{\hypertarget{ref-blei2003}{}}%
Blei, David M., Andrew Y. Ng, and Michael I. Jordan. 2003. {``Latent
Dirichlet Allocation.''} \emph{Journal of Machine Learning Research} 3
(Jan): 993--1022. \url{https://www.jmlr.org/papers/v3/blei03a}.

\leavevmode\vadjust pre{\hypertarget{ref-casella2010}{}}%
Casella, George, Malay Ghosh, Jeff Gill, and Minjung Kyung. 2010.
{``Penalized Regression, Standard Errors, and Bayesian Lassos.''}
\emph{Bayesian Analysis} 5 (2): 369411.

\leavevmode\vadjust pre{\hypertarget{ref-chakraborty2020}{}}%
Chakraborty, Saptarshi, Colin B. Begg, and Ronglai Shen. 2020. {``Using
the {"}Hidden{"} genome to improve classification of cancer types.''}
\emph{Biometrics}, September.

\leavevmode\vadjust pre{\hypertarget{ref-pkg_RcppML}{}}%
DeBruine, Zachary. 2021. \emph{{RcppML: Rcpp Machine Learning Library}}.

\leavevmode\vadjust pre{\hypertarget{ref-polson2013}{}}%
Polson, Nicholas G., James G. Scott, and Jesse Windle. 2013. {``Bayesian
Inference for Logistic Models Using {P}ólya{\textendash}Gamma Latent
Variables.''} \emph{J Am Stat Assoc} 108 (504): 1339--49.
\url{https://doi.org/10.1080/01621459.2013.829001}.

\end{CSLReferences}